\documentclass[twocolumn,showpacs]{revtex4} %
\usepackage{epsfig,amsmath,amssymb}
\usepackage{graphicx}
\usepackage[english]{babel} 
\usepackage{verbatim}

\begin{document}

\title{THERMODYNAMICS OF THE FERMI GAS IN A QUANTUM WELL}

\author{Yu.M. Poluektov}
\email{yuripoluektov@kipt.kharkov.ua} %
\affiliation{National Science Center ``Kharkov Institute of Physics and Technology'', %
Akhiezer Institute for Theoretical Physics, 61108 Kharkov, Ukraine} %
\author{A.A. Soroka} %
\affiliation{National Science Center ``Kharkov Institute of Physics and Technology'', %
Akhiezer Institute for Theoretical Physics, 61108 Kharkov, Ukraine} %

\begin{abstract}
For the ideal Fermi gas that fills a quantum well confined by two
parallel planes, there are calculated the thermodynamic
characteristics in general form for arbitrary temperatures, namely:
the thermodynamic potential, energy, entropy, equations of state,
heat capacities and compressibilities. The distance between planes
is considered as an additional thermodynamic variable. Owing to the
anisotropy, the pressure of the Fermi gas along and transverse to
the planes is different, so that the system is characterized by two
equations of state and a set of different heat capacities. Limiting
cases of low and high temperatures are considered. The temperature
dependencies of the entropy and heat capacities at low temperatures
remain linear, just as in the volume case, and their dependencies on
the chemical potential and density undergo jumps at the beginning of
the filling of new discrete levels. It is shown that the behavior of
thermodynamic quantities with the distance between plates can be
either oscillating or monotonic, depending on what quantity is
assumed to be fixed: the volume or surface density. For high
temperatures the corrections to thermodynamic quantities are
obtained, which are proportional to the ratio of the thermal de
Broglie wavelength to the distance between planes.
\newline%
{\bf Key words}: Fermi particle, quantum well, thermodynamic
functions, low-dimensional systems, equation of state, heat
capacity, compressibility
\end{abstract}
\pacs{64.10.+h, 64.60.an, 67.10.Db, 67.30.ej, 73.21.-b} 
\maketitle

\section{Introduction}
\vspace{-0mm}

The model of the ideal Fermi gas is the basis for understanding the
properties of electron and other many-fermion systems. %
In many cases it is also possible to describe with reasonable
accuracy the behavior of systems of interacting fermi-particles,
which dispersion law differs from the dispersion law of free
particles, within the approximation of an ideal gas of
quasiparticles. It is essential that thermodynamic characteristics
of the ideal Fermi gas at arbitrary temperatures in the volume case
can be expressed through the special Fermi functions and, therefore,
it is possible to obtain and verify all relations of the
phenomenological thermodynamics on the basis of the quantum
microscopic model.

In recent time, much attention has been paid to investigation of
low-dimensional systems, in particular to properties of the
two-dimensional Fermi gas in quantum wells, because apart from
purely scientific interest the study of such objects is rather
promising for the solid-state electronics \cite{Ando,Komnik,DNG,Vagner,Freik}. %
Thermodynamics relations for the Fermi gas in the confined geometry
have been studied much less than in the volume case \cite{LL} and
require further investigation. A detailed understanding of the
properties of such systems must serve as a basis for the study of
low-dimensional systems of interacting particles.

It is usual to consider that strongly correlated Fermi systems, to
which also two-dimensional Fermi liquids are attributed, in many
respects essentially differ from the usual Fermi systems and often
show ``non-Fermi-liquid'' behavior \cite{SP}. At that the properties
of quasi-two-dimensional and quasi-one-dimensional systems are
compared with the theory of bulk Fermi liquid \cite{Landau,PN}.
However, as seen even on the example of the quasi-two-dimensional
system of noninteracting particles which is considered in detail in
the present work, its properties can substantially differ,
especially at low temperatures, from the properties of the bulk
system owing to the quantum size effect. Therefore, the theory of
Fermi liquid itself in conditions of the confined geometry must,
generally speaking, be formulated differently than in the volume
case. Note that the Migdal's theory of finite Fermi systems
\cite{Migdal} does not essentially differ in this respect from the
Fermi liquid theory of uniform systems.

The consideration of low-dimensional models of interacting Fermi
particles leads to a conclusion about, in many cases, unique
properties of such systems \cite{Тоmonaga,Luttinger}. It should be
kept in mind, however, that real systems are always
three-dimensional and their low dimensionality manifests itself only
in the boundedness of motion of particles in one, two or three
coordinates. In considering statistical properties of the
three-dimensional many-particle systems one usually passes to the
thermodynamic limit, setting in final formulas the volume and number
of particles to infinity at a fixed density. It is of general
theoretical interest to study the statistical properties of
many-particle systems occupying a volume, one or two dimensions of
which remain fixed, and the thermodynamic limiting transition is
carried out only over remaining coordinates. In this case the
coordinates, over which the thermodynamic limiting transition is not
performed, should be considered as additional thermodynamic
variables. The model of the ideal Fermi gas allows to build the
thermodynamics of such systems on the basis of the statistical
treatment.

The idea of taking account of the spatial quantization when
calculating the electron heat capacity of small particles was for
the first time used by Fr\"{o}hlich \cite{Frohlich}. Thermodynamic
properties of the Fermi gas at low temperatures in the confined
geometry within the quasiclassical approach were considered in the
works of I.M. Lifshits and A.M. Kosevich \cite{LK,LK2}. Since the
fermions possess a quasidiscrete spectrum in the confined geometry,
in a similar way as it takes place for electrons in the magnetic
field \cite{LL}, the authors of these works showed that the
thermodynamic potential contains under such conditions a component
that oscillates with varying the chemical potential. However, as
seen from the results of the given work, the presence of the
oscillating component in the thermodynamic potential does not yet
guarantee that the full thermodynamic potential and thermodynamic
quantities are oscillating.

It should be noted that in experiments the quantum oscillation
phenomena in thin metallic, semimetallic, semiconducting films and
nanostructures were observed for the kinetic coefficients such as
the conductivity, the mobility, the Hall coefficient and others
\cite{Komnik,Freik}, but not for the thermodynamic quantities.

In the proposed work, on the basis of a microscopic treatment there
are obtained exact formulas for the thermodynamic potential, energy,
entropy, pressures, heat capacities and compressibilities which
allow to analyze the equilibrium properties of the Fermi system at
arbitrary temperatures and geometric dimensions. Any thermodynamic
quantity proves to be expressed through some standard functions and
their derivatives that depend only on the dimensionless temperature
and the dimensionless chemical potential. It is shown that, owing to
the anisotropy of the system under consideration, the pressure of
the Fermi gas along and transverse to planes is different, so that
the system is characterized by two equations of state and a set of
different heat capacities. The cases of low and high temperatures
are studied. At low temperatures the dependencies of the entropy and
heat capacities on temperature remain linear, just as in the volume
case, and their dependencies on the chemical potential and density
undergo jumps at the beginning of the filling of new discrete
levels. It is shown that the behavior of thermodynamic quantities
with the distance between plates can be qualitatively different,
depending on what quantity is assumed to be fixed: the volume or
surface density. Thus, at a fixed surface density the chemical
potential and pressure vary monotonically with the thickness and at
a fixed volume density these dependencies have an oscillating
character. In the area of high temperatures the corrections to
thermodynamic quantities are obtained, proportional to the ratio of
the thermal de Broglie wavelength to the distance between planes.

\vspace{-0mm}\newpage
\section{Fermi gas in a volume} %
\vspace{-0mm}

Before proceeding to consideration of the Fermi gas of particles
with mass $m$ in conditions of the confined geometry, here we give
the basic formulas for the three-dimensional Fermi gas enclosed in
the rectangular parallelepiped of volume $V=L_xL_yL_z$, lengths of
all sides of which are large. In the case when $L_i\gg 2\pi/k_i$
(${\bf k}$ is a wave vector), the wave functions of particles have
the form of plane waves\, $\varphi_{{\bf k}}({\bf
r})=V^{-1/2}e^{i{\bf k}{\bf r}}$ and the energy of particles\,
$\varepsilon_k=\hbar^2k^2/2m$. Generally, in books authors restrict
themselves to considering the two limiting cases: either the case of
high temperatures in which a description of motion of gas particles
based on the laws of the classical mechanics is valid, or the case
of very low temperatures when the Fermi gas is degenerate \cite{LL}.
Meanwhile, all thermodynamic functions of the Fermi gas at arbitrary
temperature can be exactly expressed through the special functions
\begin{equation} \label{01}
\begin{array}{l}
\displaystyle{%
  \Phi_s(t)=\frac{1}{\Gamma(s)}\int_0^{\!\infty} \frac{z^{s-1}\,dz}{e^{z-t}+1}, %
}%
\end{array}
\end{equation}
where $s$ is an integer or half-integer positive number, $\Gamma(s)$
is the gamma function. For calculation of the bulk properties of the
Fermi gas it is sufficient to know the functions (\ref{01}) with
half-integer indices $s=1/2,3/2,5/2$.

The thermodynamic potential $\Omega$, energy $E$, entropy $S$,
particle number density $n$  and pressure $p$ of the Fermi gas of
particles with the spin 1/2\, in a volume, expressed through the
functions (\ref{01}), are given by the formulas:
\begin{equation} \label{02}
\begin{array}{ll}
\displaystyle{\hspace{0mm}%
  \Omega=-\frac{2TV}{\Lambda^3}\Phi_{5/2}(t),\quad E=\frac{3TV}{\Lambda^3}\Phi_{5/2}(t), %
}\vspace{2mm}\\ %
\displaystyle{\hspace{0mm}%
  S=\frac{2V}{\Lambda^3}\left[\frac{5}{2}\Phi_{5/2}(t)-t\Phi_{3/2}(t)\right], %
}\vspace{2mm}\\ %
\displaystyle{\hspace{0mm}%
  n=\frac{N}{V}=\frac{2}{\Lambda^3}\Phi_{3/2}(t),\quad p=\frac{2T}{\Lambda^3}\Phi_{5/2}(t). %
}
\end{array}
\end{equation}
Here $t\equiv \mu/T$, $\mu$ is the chemical potential. The thermal
de Broglie wavelength enters into the formulas (\ref{02}):
\begin{equation} \label{03}
\begin{array}{l}
\displaystyle{%
  \Lambda\equiv\left(\frac{2\pi\hbar^2}{mT}\right)^{\!1/2}. %
}%
\end{array}
\end{equation}
In a bulk gas in addition to the thermal wavelength of a particle
(\ref{03}) there exists one more characteristic length $l=n^{-1/3}$
which defines an average distance between particles. The ratio of
these lengths
\begin{equation} \label{04}
\begin{array}{l}
\displaystyle{%
  q_\Lambda\equiv\frac{\Lambda}{l}=\Lambda n^{1/3}= \left[2\Phi_{3/2}(t)\right]^{1/3} %
}%
\end{array}
\end{equation}
characterizes the extent of proximity of the gas to the degenerate
state, so to say the measure of its ``quantumness''. Depending on
the density of number of particles and temperature the quantum
mechanical properties of the gas will manifest themselves to a
greater or lesser extent. If $q_\Lambda$ is small, the system can be
well described by the classical mechanics. With decreasing
temperature at a fixed density the thermal wavelength increases and,
therefore, the ``quantumness'' of the gas increases in this case,
and besides, as we see, the parameter $q_\Lambda$ depends on the
single parameter $t$.

Here we also give the formulas for the heat capacities at a constant
volume and a constant pressure:
\begin{equation} \label{05}
\begin{array}{ll}
\displaystyle{%
  C_V=\frac{15}{2}\frac{V}{\Lambda^3}\!\left[ \Phi_{5/2}(t) - \frac{3}{5}\frac{\Phi_{3/2}^2(t)}{\Phi_{1/2}(t)} \right], %
}\vspace{2mm}\\ %
\displaystyle{\hspace{0mm}%
  C_p=\frac{25}{2}\frac{V}{\Lambda^3}\Phi_{5/2}(t)\!\left[ \frac{\Phi_{1/2}(t)\Phi_{5/2}(t)}{\Phi_{3/2}^2(t)} -\frac{3}{5} \right]. %
}
\end{array}
\end{equation}
It should be noted that the heat capacities per one particle
$C_V\big/N$ and $C_p\big/N$ also depend on the single parameter $t$.
The properties of Fermi systems in the presence of the discrete
levels in magnetic field were studied with the help of the functions
(\ref{01}) in work \cite{Poluektov}.

\section{Thermodynamics of the Fermi gas in a rectangular quantum well} %
The model of the ideal Fermi gas is the basis for studying the bulk
properties of the Fermi systems for particles of different nature.
In the two-dimensional case an analogous role is played by the ideal
Fermi gas contained between two parallel planes, therefore a
detailed study of such the system is also of general physical
interest. In particular, it is important to obtain exact formulas
for thermodynamic quantities of the Fermi gas contained between two
parallel planes $z=L/2$, $z=-L/2$ and to make analysis of its
thermodynamic properties. It is assumed everywhere that the spin of
the Fermi particle is equal to 1/2. The lengths $L_x, L_y$ are as
before considered to be macroscopic, where $A=L_xL_y$ is the area in
the $(x,y)$ plane, but no restrictions are imposed on the length of
the third side $L_z=L$ and it can be small, that corresponds to
transition to the quasi-two-dimensional case. The total volume
occupied by the Fermi gas $V=AL$. The case when $L_x\gg L$, $L_y\gg
L$ is of the most interest and the main attention will be paid to
it. Note that in work \cite{LK2} this case was not considered in
detail. Let us assume that the potential barrier at the points
$z=L/2$ and $z=-L/2$ is infinite, so that the wave function of a
particle turns into zero at boundaries. In this case solutions of
the Schr\"{o}dinger equation have the form
\begin{equation} \label{06}
\begin{array}{ll}
\displaystyle{%
  \varphi_{{\bf k},n}^{(+)}(x,y,z)=\!\sqrt{\frac{2}{AL}}\,e^{i{\bf k}{\bf r}}\!\cos(2n+1)\frac{\pi z}{L},\,\, (n=0,1,\ldots), %
}\vspace{2mm}\\ %
\displaystyle{\hspace{0mm}%
  \varphi_{{\bf k},n}^{(-)}(x,y,z)=\!\sqrt{\frac{2}{AL}}\,e^{i{\bf k}{\bf r}}\!\sin2n\frac{\pi z}{L},\,\, (n=1,2,\ldots). %
}
\end{array}
\end{equation}
The first of these functions is even and the second is odd with
respect to the transformation $z\rightarrow -z$. Here ${\bf k}\equiv
(k_x,k_y)$ and ${\bf r}\equiv (x,y)$ are two-dimensional vectors.
The energy of a particle:
\begin{equation} \label{07}
\begin{array}{ll}
\displaystyle{%
  \varepsilon_{kn}=\frac{\hbar^2k^2}{2m}+\varepsilon_L n^2, %
}
\end{array}
\end{equation}
$n=1,2,\ldots,$ and $k^2=k_x^2+k_y^2$. For odd $n$ the levels
correspond to even wave functions, and for even $n$ -- to odd wave
functions. The energy
\begin{equation} \label{08}
\begin{array}{ll}
\displaystyle{%
  \varepsilon_L\equiv\frac{\pi^2\hbar^2}{2mL^2} %
}
\end{array}
\end{equation}
is conditioned by localization of a quantum particle between planes
and increases with decreasing the distance $L$ between them. This
characteristic energy of a problem (we assume the particle is an
electron), expressed in Rydbergs, can be written in the form
$\tilde{\varepsilon}_L=\pi^2\big/\tilde{L}^2$, where
$\tilde{L}=L/a_0$ is the distance in the Bohr radiuses. %
Considering that $1\,\rm{Ry}=13.6$\,eV\,$=15.8\cdot 10^4$\,K, we obtain: %
1) at $\tilde{L}=10^2$ -- $\tilde{\varepsilon}_L=156$\,K; %
2) at $\tilde{L}=10^3$ -- $\tilde{\varepsilon}_L=1.56$\,K; %
3) at $\tilde{L}=10^4$ -- $\tilde{\varepsilon}_L=0.016$\,K; %
4) at $\tilde{L}=10^6$ -- $\tilde{\varepsilon}_L=1.6\cdot 10^{-6}$\,K. %
Thus, quantum effects connected with the presence of the energy
(\ref{08}), as it had to be expected, must manifest themselves
essentially at low temperatures and small distances between planes. %

The distribution function in this case has the form
\begin{equation} \label{09}
\begin{array}{ll}
\displaystyle{%
  f_{kn}=\big[\exp(\varepsilon_{kn}-\mu)\big/T +1\big]^{-1}.  %
}
\end{array}
\end{equation}
After integration over momenta with the function (\ref{09}), the
thermodynamical potential
\begin{equation} \label{10}
\begin{array}{ll}
\displaystyle{%
  \Omega=-2T\sum_{k,n}\ln\!\left[1+e^{-(\varepsilon_{kn}-\mu)/T}\right] %
}
\end{array}
\end{equation}
will be determined by the formula
\begin{equation} \label{11}
\begin{array}{ll}
\displaystyle{%
  \Omega=-\frac{2TA}{\Lambda^2}\Psi_2(\tau,\eta). %
}
\end{array}
\end{equation}
Instead of the parameter $t\equiv \mu/T$, that was used in formulas
for the volume case, it is convenient to introduce the dimensionless
chemical potential $\eta\equiv \mu/\varepsilon_L$ and the
dimensionless temperature $\tau\equiv T/\varepsilon_L$, then
$t=\eta/\tau$. We define the function
\begin{equation} \label{12}
\begin{array}{ll}
\displaystyle{%
  \Psi_s(\tau,\eta)\equiv\sum_{n=1}^\infty \Phi_s \Big[\tau^{-1}\big(\eta-n^2\big)\Big]. %
}
\end{array}
\end{equation}
The details of calculation of such functions are given in Appendix\,\ref{A}. %
The thermodynamic potential (\ref{11}) is a function of the
temperature, chemical potential, area and distance between plates:
$\Omega=\Omega(T,\mu,A,L)$. In contrast to the volume case when
$\Omega$ is proportional to the volume $V$, in this case it is
proportional to the area $A$ and depends in a complicated manner on
the distance $L$. This circumstance is conditioned by the evident
anisotropy of the system under consideration, since here the motions
in the $(x,y)$ plane and in the direction of the $z$ axis are
qualitatively different. In statistical mechanics it is customary to
pass in the final formulas to the thermodynamic limit
$V\rightarrow\infty$, $N\rightarrow\infty$ at $n=N/V=\rm{const}$. In
the present case it is more accurate to write down the thermodynamic
limit somewhat differently, namely
\begin{equation} \label{13}
\begin{array}{ll}
\displaystyle{%
  A\rightarrow\infty,\, N\rightarrow\infty \textrm{\,\,\,\,\,at\,\,\,\,} n_{\!A}\equiv N/A=\rm{const}. %
}
\end{array}
\end{equation}
It is thereby stressed that the transition to infinite volume occurs
only owing to increasing the area, at a fixed distance $L$.

The differential of the thermodynamic potential (\ref{11}) has the form %
\begin{equation} \label{14}
\begin{array}{ll}
\displaystyle{%
  d\Omega= -\frac{2A}{\Lambda^2}\!\left( 2\Psi_2 + \tau\frac{\partial\Psi_2}{\partial\tau} \right)\!dT - \frac{2A}{\Lambda^2}\Psi_1d\mu + \frac{\Omega}{A}dA\,-%
}\vspace{2mm}\\ %
\displaystyle{\hspace{-0mm}%
  \hspace{09mm}-\frac{4AT}{\Lambda^2L}\!\left( \tau\frac{\partial\Psi_2}{\partial\tau}+ \eta\frac{\partial\Psi_2}{\partial\eta} \right)\!dL. %
}
\end{array}
\end{equation}
It was taken into account that %
$\partial\Psi_2\big/\partial\eta=\tau^{-1}\Psi_1$ and $d\varepsilon_L=-(2\varepsilon_L/L)dL$. %
Since $S=-\big(\partial\Omega\big/\partial T\big)_{\mu,A,L}$ and $N=-\big(\partial\Omega\big/\partial \mu\big)_{T,A,L}$, %
from (\ref{14}) there follow expressions for the entropy and number of particles: %
\begin{equation} \label{15}
\begin{array}{ll}
\displaystyle{%
  S=\frac{2A}{\Lambda^2}\!\left( 2\Psi_2 + \tau\frac{\partial\Psi_2}{\partial\tau} \right), %
}
\end{array}
\end{equation}
\begin{equation} \label{16}
\begin{array}{ll}
\displaystyle{%
  N=\frac{2A}{\Lambda^2}\Psi_1 . %
}
\end{array}
\end{equation}
The volume and surface densities of number of particles are defined
by obvious relations: $n\equiv N/AL$, $n_{\!A}\equiv N/A$. The same
formulas for the entropy and number of particles can be derived, of
course,  straight by means of the distribution function (\ref{09}).
The energy is determined by the formula:
\begin{equation} \label{17}
\begin{array}{ll}
\displaystyle{%
  E=\frac{2AT}{\Lambda^2}\!\left( \Psi_2 + \frac{\eta}{\tau}\Psi_1 + \tau\frac{\partial\Psi_2}{\partial\tau} \right). %
}
\end{array}
\end{equation}
Naturally, the relation holds $\Omega=E-TS-\mu N$.

\section{Pressures}
In a bulk system the pressure is connected with the thermodynamical
potential by the known formula $p=-\Omega/V$. In the considered case
the system is anisotropic, since the character of motion of
particles in the directions parallel and perpendicular to planes is
different, and, therefore, the usual formula for the pressure is
invalid. The force exerted by the gas on the wall perpendicular to
the $z$ axis is different from the force exerted on the side walls
perpendicular to the $x$ and $y$ axes. These forces can be
calculated in the same way as in the volume case \cite{LL}. The
pressures in directions parallel to the $(x,y)$ plane and on the
planes perpendicular to the $z$ axis are given by the formulas
\begin{equation} \label{18}
\begin{array}{ll}
\displaystyle{%
  p_\parallel =-\frac{1}{L}\!\left( \frac{\partial E}{\partial A}\right)_{\!S,L}, \quad  %
  p_\perp =-\frac{1}{A}\!\left( \frac{\partial E}{\partial L}\right)_{\!S,A} .   %
}
\end{array}
\end{equation}
Since
\begin{equation} \label{19}
\begin{array}{ll}
\displaystyle{%
  \left( \frac{\partial E}{\partial A}\right)_{\!S,N,L}\!=\!\left( \frac{\partial \Omega}{\partial A}\right)_{\!T,\mu,L}\!, \quad  %
  \left( \frac{\partial E}{\partial L}\right)_{\!S,N,A}\!=\!\left( \frac{\partial \Omega}{\partial L}\right)_{\!T,\mu,A}\!,  %
}
\end{array}
\end{equation}
it is then more convenient to calculate the pressures (\ref{18})
using the formulas
\begin{equation} \label{20}
\begin{array}{ll}
\displaystyle{%
  p_\parallel =-\frac{1}{L}\!\left( \frac{\partial \Omega}{\partial A}\right)_{\!T,\mu,L}, \quad  %
  p_\perp =-\frac{1}{A}\!\left( \frac{\partial \Omega}{\partial L}\right)_{\!T,\mu,A} .   %
}
\end{array}
\end{equation}
The differential of the thermodynamic potential (\ref{14}) can be
represented in the form
\begin{equation} \label{21}
\begin{array}{ll}
\displaystyle{%
  d\Omega=-SdT-Nd\mu-p_\parallel LdA-p_\perp AdL .   %
}
\end{array}
\end{equation}
Considering the form of the thermodynamic potential (\ref{11}), we
obtain the formulas determining the pressures through the functions
(\ref{12}):
\begin{equation} \label{22}
\begin{array}{ll}
\displaystyle{%
  p_\parallel =\frac{2T}{\Lambda^2L}\Psi_2, \quad  %
  p_\perp =\frac{4T}{\Lambda^2L}\!\left( \tau\frac{\partial\Psi_2}{\partial\tau}+ \frac{\eta}{\tau}\Psi_1 \right) .   %
}
\end{array}
\end{equation}
The quantity $-p_\parallel L$ is an analog of the surface tension in
the theory of surfaces \cite{LL}. The energy (\ref{17}) is connected
with the pressures (\ref{22}) by the relation
\begin{equation} \label{23}
\begin{array}{ll}
\displaystyle{%
  E=AL \!\left( p_\parallel+\frac{1}{2}\,p_\perp \right),  %
}
\end{array}
\end{equation}
which in the volume limit $p_\parallel=p_\perp=p$ \,turns into the
known relation $pV=(2/3)E$ \,for the Fermi gas \cite{LL}.

\section{Reduced form of thermodynamic quantities} %
It is convenient to introduce dimensionless quantities, which we
will call ``reduced'' and designate them by a tilde on top, for the
entropy, energy, pressures, volume and surface densities:
\begin{equation} \label{24}
\begin{array}{ll}
\displaystyle{%
  \tilde{S}\equiv\frac{2L^2}{\pi A}S, \qquad \tilde{E}\equiv\frac{2mL^4}{\pi^3\hbar^2A}E, %
}\vspace{2mm}\\ %
\displaystyle{\hspace{0mm}%
  \tilde{p}_\parallel\equiv\frac{2mL^5}{\pi^3\hbar^2}p_\parallel, \qquad \tilde{p}_\perp\equiv\frac{2mL^5}{\pi^3\hbar^2}p_\perp , %
}\vspace{2mm}\\ %
\displaystyle{\hspace{0mm}%
\tilde{n}\equiv\frac{2L^3}{\pi}n, \qquad \tilde{n}_{\!A}\equiv\frac{2L^2}{\pi}n_{\!A}. %
}
\end{array}
\end{equation}
The reduced quantities are functions of only two independent
dimensionless variables -- the temperature $\tau$ and chemical
potential $\eta$: 
\begin{equation} \label{25}
\begin{array}{ll}
\displaystyle{%
   \tilde{S}=\tau\!\left( 2\Psi_2+\tau\frac{\partial\Psi_2}{\partial\tau} \right),  %
}\vspace{2mm}\\ %
\displaystyle{\hspace{0mm}%
  \tilde{E}=\frac{\tau^2}{2}\!\left( \Psi_2+\frac{\eta}{\tau}\Psi_1+\tau\frac{\partial\Psi_2}{\partial\tau} \right), %
}\vspace{2mm}\\ %
\displaystyle{\hspace{0mm}%
  \tilde{n}=\tilde{n}_{\!A}=\tau\Psi_1, %
}\vspace{2mm}\\ %
\displaystyle{\hspace{0mm}%
  \tilde{p}_\parallel=\frac{\tau^2}{2}\Psi_2,\quad \tilde{p}_\perp=\tau^2\!\left(\frac{\eta}{\tau}\Psi_1+\tau\frac{\partial\Psi_2}{\partial\tau} \right). %
}
\end{array}
\end{equation}
The use of the reduces quantities is convenient owing to the fact
that they do not contain explicitly geometric dimensions of the
system.

\section{Heat capacities}\vspace{-0mm}
An important directly observable thermodynamic quantity is the heat
capacity. In the geometry under consideration heat capacities can be
defined under various conditions different from that which take
place in the volume case. In order to determine heat capacities, it
is necessary to calculate the quantity $C=T(dS/dT)$. For this
purpose, it is convenient to express the differential of the entropy
through the reduced quantities:
\begin{equation} \label{26}
\begin{array}{ll}
\displaystyle{\hspace{-3mm}%
   dS=\frac{\pi A}{2L^2}\!\left( -\frac{\tilde{S}}{\tilde{n}_{\!A}}\,d\tilde{n}_{\!A} + d\tilde{S} \right)=  %
}\vspace{2mm}\\ %
\displaystyle{\hspace{-1mm}%
  =\frac{\pi A}{2L^2}\!
  \left[\!
  \left(\frac{\partial \tilde{S}}{\partial\eta}-\frac{\tilde{S}}{\tilde{n}_{\!A}}\frac{\partial \tilde{n}_{\!A}}{\partial\eta}\right)\!d\eta + \! %
  \left(\frac{\partial \tilde{S}}{\partial\tau}-\frac{\tilde{S}}{\tilde{n}_{\!A}}\frac{\partial \tilde{n}_{\!A}}{\partial\tau}\right)\!d\tau
  \right]\!. %
}
\end{array}
\end{equation}
In the volume case at a fixed number of particles, which is assumed
here, the equation of state is: $p=p(T,V)$. If the chemical
potential is used as an independent variable, then the equation of
state is defined parametrically by the equations $p=p(T,V,\mu)$ and
$N=N(T,V,\mu)$. To obtain the heat capacity as a function of only
temperature, one constraint should be imposed between the pressure
and the volume. In the simplest case, it its possible to fix either
the volume or the pressure, thus determining the heat capacities
$C_V$ and $C_p$.

Under given conditions, owing to anisotropy of the system, there are
two equations of state (\ref{22}) for two pressures
$p_\parallel=p_\parallel(T,A,L,\mu)$ and $p_\perp=p_\perp(T,A,L,\mu)$, %
which at a fixed number of particles should be considered together
with the equation (\ref{16}) $N=N(T,A,L,\mu)$. To obtain the heat
capacity as a function of only temperature, two additional
constraints should be set between the pressures $p_\parallel$,
$p_\perp$ and the dimensions of the system $A,L$, namely
$F_1(p_\parallel,p_\perp,A,L)=0$ and $F_2(p_\parallel,p_\perp,A,L)=0$. %
In the simplest case, two of four quantities
$p_\parallel,p_\perp,A,L$ can be fixed. %
Then the heat capacity as a function of temperature can be
considered under fixation of one of the following pairs of
quantities: $(A,L)$, $(p_\parallel,p_\perp)$, $(A, p_\parallel)$,
$(A, p_\perp)$, $(L, p_\parallel)$, $(L, p_\perp)$. Fixation of the
first of pairs $(A,L)$ corresponds to the volume case of the heat
capacity at a constant volume, and of the second
$(p_\parallel,p_\perp)$ -- at a constant pressure.

With account of the fixation of a number of particles, we have 
\begin{equation} \label{27}
\begin{array}{ll}
\displaystyle{%
   d\tilde{n}_{\!A}=\tilde{n}_{\!A}\!\left( 2\frac{dL}{L}-\frac{dA}{A} \right).  %
}
\end{array}
\end{equation}
Also it should be taken into account that %
\begin{equation} \label{28}
\begin{array}{ll}
\vspace{5mm}
\displaystyle{%
   \frac{d\tau}{dT}=\frac{1}{\varepsilon_L}+\frac{2\tau}{L}\frac{dL}{dT}.  \vspace{0mm} %
}
\end{array}
\end{equation}
Finally, we obtain the formulas for the reduced heat capacities
$\displaystyle{\tilde{C}\equiv\frac{2L^2}{\pi A}\,C}$ under
different conditions, being valid at arbitrary temperatures:
\widetext\vspace{-0mm}
\begin{equation} \label{29} %
\begin{array}{ll}
\displaystyle{%
   \tilde{C}_{AL}=\tau\!\left\{ \frac{\partial \tilde{S}}{\partial\tau}- %
   \frac{\partial \tilde{S}}{\partial\eta}\frac{\big(\partial\tilde{n}_{\!A}/\partial\tau\big)}{\big(\partial\tilde{n}_{\!A}/\partial\eta\big)} \right\},  %
}
\end{array}
\end{equation}
\begin{equation} \label{30}
\begin{array}{ll}
\displaystyle{%
   \tilde{C}_{p_\parallel p_\perp}=\tau\,
   \frac{
   \displaystyle{
   \left(\frac{\partial \tilde{S}}{\partial\eta}-\frac{\tilde{S}}{\tilde{n}_{\!A}}\frac{\partial \tilde{n}_{\!A}}{\partial\eta}\right)\!\!
   \left(\tilde{p}_\perp\frac{\partial \tilde{p}_\parallel}{\partial\tau}-\tilde{p}_\parallel\frac{\partial \tilde{p}_\perp}{\partial\tau}\right) + \! %
   \left(\frac{\partial \tilde{S}}{\partial\tau}-\frac{\tilde{S}}{\tilde{n}_{\!A}}\frac{\partial \tilde{n}_{\!A}}{\partial\tau}\right)\!\!
   \left(\tilde{p}_\parallel\frac{\partial \tilde{p}_\perp}{\partial\eta}-\tilde{p}_\perp\frac{\partial \tilde{p}_\parallel}{\partial\eta}\right) %
   }
   }
   {
   \displaystyle{
   \tilde{p}_\parallel\frac{\partial \tilde{p}_\perp}{\partial\eta}-\tilde{p}_\perp\frac{\partial \tilde{p}_\parallel}{\partial\eta} %
   +\frac{2}{5}\,\tau\!\left(\frac{\partial \tilde{p}_\parallel}{\partial\eta}\frac{\partial \tilde{p}_\perp}{\partial\tau}- %
   \frac{\partial \tilde{p}_\parallel}{\partial\tau}\frac{\partial \tilde{p}_\perp}{\partial\eta}\right) %
   }
   },
}
\end{array}
\end{equation}
\begin{equation} \label{31}
\begin{array}{ll}
\displaystyle{%
   \tilde{C}_{L p_\parallel}=\tau\!\left\{\!
   \left(\frac{\partial \tilde{S}}{\partial\tau}-\frac{\tilde{S}}{\tilde{n}_{\!A}}\frac{\partial \tilde{n}_{\!A}}{\partial\tau}\right)-\! %
   \left(\frac{\partial \tilde{S}}{\partial\eta}-\frac{\tilde{S}}{\tilde{n}_{\!A}}\frac{\partial \tilde{n}_{\!A}}{\partial\eta}\right)\!\! %
   \frac{\big(\partial\tilde{p}_\parallel/\partial\tau\big)}{\big(\partial\tilde{p}_\parallel/\partial\eta\big)} \right\},  %
}
\end{array}
\end{equation}
\begin{equation} \label{32}
\begin{array}{ll}
\displaystyle{%
   \tilde{C}_{A p_\parallel}=\tau\,
   \frac{
   \displaystyle{
   \tilde{n}_{\!A}\frac{\partial \tilde{p}_\parallel}{\partial\eta}\!
   \left(\frac{\partial \tilde{S}}{\partial\tau}-\frac{\tilde{S}}{\tilde{n}_{\!A}}\frac{\partial \tilde{n}_{\!A}}{\partial\tau}\right)\!- %
   \tilde{n}_{\!A}\frac{\partial \tilde{p}_\parallel}{\partial\tau}\!
   \left(\frac{\partial \tilde{S}}{\partial\eta}-\frac{\tilde{S}}{\tilde{n}_{\!A}}\frac{\partial \tilde{n}_{\!A}}{\partial\eta}\right)\!+ %
   \frac{5}{2}\,\tilde{p}_\parallel\!
   \left(\frac{\partial \tilde{S}}{\partial\eta}\frac{\partial \tilde{n}_{\!A}}{\partial\tau}-\frac{\partial \tilde{S}}{\partial\tau}\frac{\partial \tilde{n}_{\!A}}{\partial\eta}\right) %
   }
   }
   {
   \displaystyle{
   \tilde{n}_{\!A}\frac{\partial \tilde{p}_\parallel}{\partial\eta}-\frac{5}{2}\,\tilde{p}_\parallel\frac{\partial \tilde{n}_{\!A}}{\partial\eta}+ %
   \tau\!\left(\frac{\partial \tilde{n}_{\!A}}{\partial\eta}\frac{\partial \tilde{p}_\parallel}{\partial\tau}-\frac{\partial \tilde{n}_{\!A}}{\partial\tau}\frac{\partial \tilde{p}_\parallel}{\partial\eta}\right) %
   }
   }.
}
\end{array}
\end{equation}
\endwidetext
The heat capacities $\tilde{C}_{L p_\perp},\tilde{C}_{A p_\perp}$ %
are determined by formulas (\ref{31}),(\ref{32}) 
with account of the replacement $p_\parallel\rightarrow p_\perp.\!\!$ 

\vspace{0mm} %
\begin{figure*}[t!]
\vspace{0mm} \hspace{-0mm}
\includegraphics[width = 1.0\textwidth]{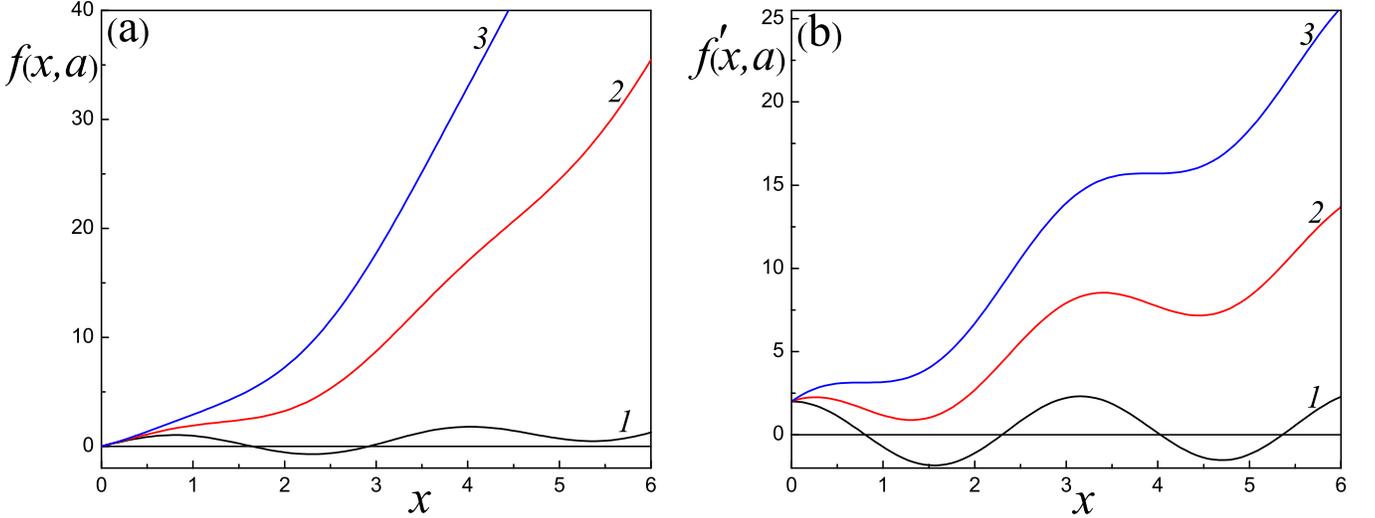} 
\vspace{-7mm}
\caption{\label{fig01} %
(a) Graphs of the function $f(x,a)=\frac{1}{2}ax^2 + \sin 2x$ for the values of the parameter $a$: %
({\it 1}) 0.1; ({\it 2}) 2.0; ({\it 3}) 4.0. \newline%
(b) Graphs of the derivative $f'(x,a)=ax + 2\cos 2x$ for the same values of the parameter $a$: %
({\it 1}) 0.1; ({\it 2}) 2.0; ({\it 3}) 4.0.
}%
\end{figure*}

\section{Compressibilities}
Another directly observable quantities are compressibilities. We
define ``parallel'' and ``perpendicular'' compressibilities by the
relations
\begin{equation} \label{33}
\begin{array}{ll}
\displaystyle{%
   \gamma_\parallel = \frac{1}{n}\!\left( \frac{\partial n}{\partial p_\parallel}\right)_{\!L},\qquad  %
   \gamma_\perp = \frac{1}{n}\!\left( \frac{\partial n}{\partial p_\perp}\right)_{\!A}.  %
}
\end{array}
\end{equation}
Compressibilities can be calculated under condition of constant
temperature (isothermal) and constant entropy (adiabatic). For
compressibilities in isothermal conditions, defined by the relations
(\ref{33}), we obtain:
\begin{equation} \label{34}
\begin{array}{ll}
\displaystyle{%
   \tilde{\gamma}_{\parallel T} \equiv \frac{\pi^3\hbar^2}{4mL^5}\,\gamma_{\parallel T}= %
   \frac{1}{2\tilde{n}}\frac{\big(\partial \tilde{n}/\partial\eta\big)}{\big(\partial\tilde{p}_\parallel/\partial\eta\big)}, %
}
\end{array}
\end{equation}
\begin{equation} \label{35}
\begin{array}{ll}
\displaystyle{%
   \tilde{\gamma}_{\perp T} \equiv \frac{\pi^3\hbar^2}{4mL^5}\,\gamma_{\perp T}= %
}\vspace{2mm}\\ %
\displaystyle{\hspace{0mm}%
  =\frac{\big(\partial \tilde{n}/\partial\eta\big)}
  {\displaystyle{
  4\!\left[\left(\frac{5}{2}\,\tilde{p}_\perp-\tau\frac{\partial \tilde{p}_\perp}{\partial\tau}\right)\!\!\frac{\partial\tilde{n}}{\partial\eta} -\! %
  \left(\tilde{n}-\tau\frac{\partial \tilde{n}}{\partial\tau}\right)\!\!\frac{\partial \tilde{p}_\perp}{\partial\eta} %
  \right]} %
  }.
}
\end{array}
\end{equation}

The adiabaticity condition consists in the invariance of the entropy
per one particle (and therefore of the total entropy in the system
with a fixed number of particles). In the volume case in adiabatic
processes the parameter $t=\mu/T$ is constant. It is easy to verify,
using the formulas (\ref{02}), that this condition leads to the
known equations of the adiabat: $n\big/T^{3/2}=C_1$,
$p\big/n^{5/3}=C_2$, $p\big/T^{5/2}=C_3$, where $C_i$ are constants.

In the considered case the adiabaticity condition has the form:
\begin{equation} \label{36}
\begin{array}{ll}
\displaystyle{%
   \sigma\equiv \frac{S}{N}= \frac{1}{\Psi_1}\!\left( 2\Psi_2+\tau\frac{\partial\Psi_2}{\partial\tau} \right)\!\equiv  %
   \Theta(\tau,\eta)=\rm{const}.
}
\end{array}
\end{equation}

Together with the equation for the number of particles (\ref{16})
the equation (\ref{36}) determines relationships between the
density, temperature and pressures in adiabatic processes. The
adiabatic compressibilities are given by the formulas:
\begin{equation} \label{37}
\begin{array}{ll}
\displaystyle{%
\tilde{\gamma}_{\parallel \sigma} = %
  \frac{
  \displaystyle{
  \frac{\partial\Theta}{\partial\eta}\frac{\partial\tilde{n}}{\partial\tau} - \frac{\partial\Theta}{\partial\tau}\frac{\partial\tilde{n}}{\partial\eta} %
  }
  }
  {\displaystyle{
  2\tilde{n}\!\left[\frac{\partial\Theta}{\partial\eta}\frac{\partial\tilde{p}_\parallel}{\partial\tau} - \frac{\partial\Theta}{\partial\tau}\frac{\partial\tilde{p}_\parallel}{\partial\eta}\right]} %
  },
}
\end{array}
\end{equation}
\begin{equation} \label{38}
\begin{array}{ll}
\displaystyle{%
\tilde{\gamma}_{\perp \sigma} = %
}\vspace{2mm}\\ %
\displaystyle{\hspace{0mm}%
  =\!\frac{
  \displaystyle{
  \frac{\partial\Theta}{\partial\eta}\frac{\partial\tilde{n}}{\partial\tau} - \frac{\partial\Theta}{\partial\tau}\frac{\partial\tilde{n}}{\partial\eta} %
  }
  }
  {\displaystyle{
  4\!\left[\frac{5\tilde{p}_\perp}{2}\!\left(\!\frac{\partial\Theta}{\partial\eta}\frac{\partial \tilde{n}}{\partial\tau} -\! %
  \frac{\partial\Theta}{\partial\tau}\frac{\partial\tilde{n}}{\partial\eta}\!\right)\! -
  \tilde{n}\!\left(\!\frac{\partial\Theta}{\partial\eta}\frac{\partial\tilde{p}_\perp}{\partial\tau} -\! %
  \frac{\partial\Theta}{\partial\tau}\frac{\partial\tilde{p}_\perp}{\partial\eta}\!\right)
  \right]} %
  }.
}
\end{array}
\end{equation}
Certainly, at zero temperature the isothermal and adiabatic
compressibilities coincide.

\vspace{0mm} %
\begin{figure}[t!]
\vspace{0mm} \hspace{-0mm}
\includegraphics[width = 1.0\columnwidth]{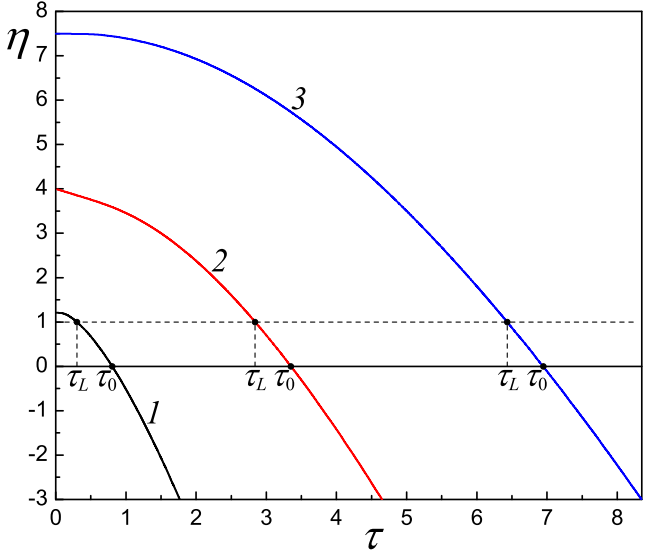} 
\vspace{-8mm}
\caption{\label{fig02} %
The dependencies of the chemical potential on temperature
$\eta(\tau)$ at different values of the reduced density:
({\it 1}) $\tilde{n}_{\!A}=0.21$, $\tau_L=0.30$, $\tau_0=0.81$; %
({\it 2}) $\tilde{n}_{\!A}=3.0$,  $\tau_L=2.84$, $\tau_0=3.35$; %
({\it 3}) $\tilde{n}_{\!A}=10.0$, $\tau_L=6.43$, $\tau_0=6.95$. %
}%
\end{figure}

\section{Analysis of functions \newline $\Psi_1(\tau,\eta)$ and $\Psi_2(\tau,\eta)$}
As shown above, all thermodynamic quantities are expressed through
the functions $\Psi_1(\tau,\eta)$, $\Psi_2(\tau,\eta)$ and their
derivatives. In this section we study the properties of these
functions. The details of calculations are given in Appendix\,\ref{A}. %
Note that when studying oscillations in the Fermi gas with quantized
levels, usually the Poisson formula is used for the extraction of an
oscillating part \cite{LL,LK,LK2}. But a detailed analysis
undertaken by the authors shows that it is more convenient to
calculate the standard functions (\ref{12}), by
which thermodynamic quantities are expressed, without use of the Poisson formula. %
This, in particular, is connected with the fact that the possibility
of extraction of an oscillating part in some function does not at
all mean that the total function is oscillating, and the
contribution of non-oscillating part should be analyzed as well. As
a simples example let us consider the function
$f(x,a)=\frac{1}{2}ax^2 + \sin 2x$. %
Despite this function contains an oscillating term, its behavior
depends on the value of its non-oscillating part, that is, the value
of the parameter $a$. The form of this function and its derivative
at some values of $a$ is shown in Fig.\,1. At $a=0.1$ both the
function and its derivative oscillate (curves {\it 1}). At $a=2$ the
function itself already proves to be monotonically increasing, while
its derivative remains oscillating (curves {\it 2}). And at $a=4$
both the function and its derivative monotonically increase (curves
{\it 3}). As it will be seen, a similar situation takes place as
well for the functions considered in the present work. Also it
should be noted that for establishing correct thermodynamic
relations, the total thermodynamic potential should be considered,
with account of contributions of both oscillating and
non-oscillating parts.

At fixed particle number density and at high temperatures, the same
as in the volume case, the chemical potential is negative. With
decreasing temperature it increases and at some temperature $T_0$
turns into zero $(\eta=0)$, becoming further positive. There is one
more characteristic temperature $T_L$, at which $\mu=\varepsilon_L$
$(\eta=1)$. The dependencies of the dimensionless chemical potential
$\eta$ on the dimensionless temperature $\tau$ are shown in Fig.\,2.
The characteristic temperatures $\tau_0=T_0/\varepsilon_L$ and
$\tau_L=T_L/\varepsilon_L$ are determined from the equations:
\begin{equation} \label{39}
\begin{array}{ll}
\displaystyle{%
   \tilde{n}_{\!A}=\tau_0\Psi_1(\tau_0,0),\quad \tilde{n}_{\!A}=\tau_L\Psi_1(\tau_L,1). %
}
\end{array}
\end{equation}
The region where $\mu\leq\varepsilon_L$ $(\eta\leq 1)$ will be for
convenience called the high temperature region, and the region
$\mu>\varepsilon_L$ $(\eta> 1)$ -- the low temperature region. The
functions $\Psi_s(\tau,\eta)$ are calculated differently in these
regions. At $\eta\leq 1$ they can be calculated by the formula
\begin{equation} \label{40}
\begin{array}{ll}
\displaystyle{%
   \Psi_s(\tau,\eta)=\frac{1}{2}\sum_{l=1}^\infty\!
   \frac{(-1)^{l+1}}{l^s}\,e^{l\eta/\tau}\theta_3\!\!\left(0,\frac{l}{\pi^2\tau}\right)\! %
   -\frac{1}{2}\,\Phi_s\left(\frac{\eta}{\tau}\right), %
}
\end{array}
\end{equation}
where $\theta_3(\nu,x)\equiv 1+2\sum_{k=1}^\infty  e^{-k^2\pi^2 x}\cos 2\pi\nu x$\, is the theta-function. %

\vspace{0mm} %
\begin{figure*}[t!]
\vspace{0mm} \hspace{-0mm}
\includegraphics[width = 1.0\textwidth]{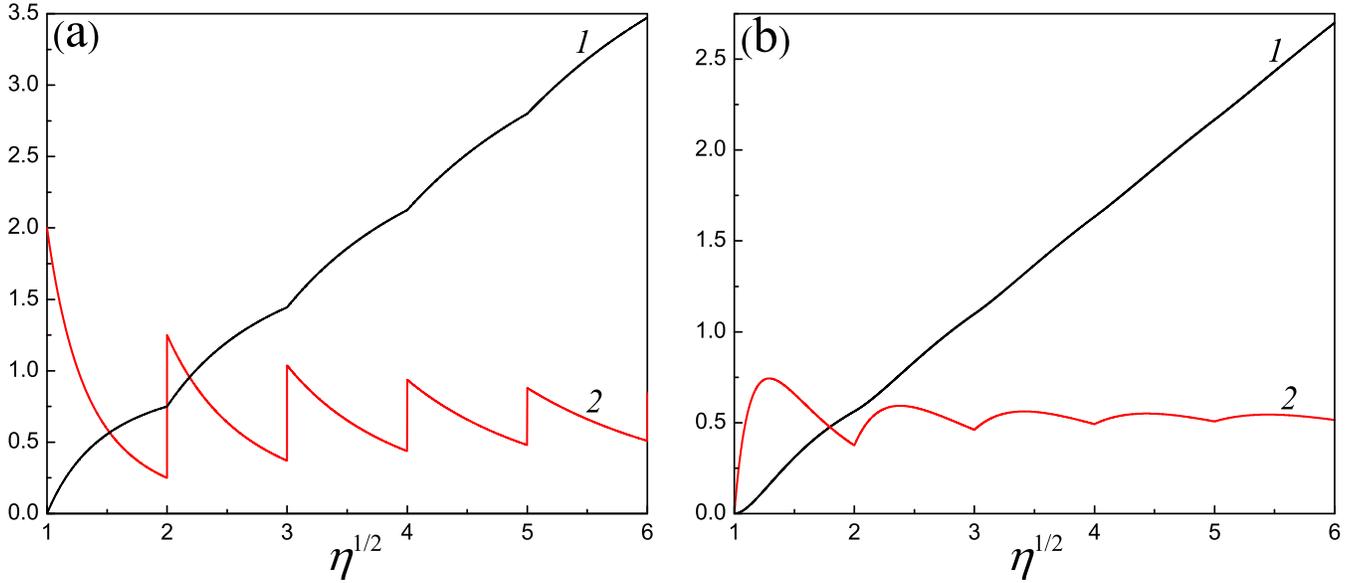} 
\vspace{-6mm}
\caption{\label{fig03} %
Graphs of the functions $\Psi_1'(\eta)$, $\Psi_2'(\eta)$ and their derivatives. \newline %
(a) The functions $\Psi_1'(\eta)$  ({\it 1}) and $d\Psi_1'(\eta)\big/d\!\sqrt{\eta}$ \,({\it 2}); %
(b) The functions $\Psi_2'(\eta)$  ({\it 1}) and $d\Psi_2'(\eta)\big/d\!\sqrt{\eta}$ \,({\it 2}). %
}%
\end{figure*}

\vspace{0mm} %
\begin{figure*}[t!]
\vspace{0mm} \hspace{-0mm}
\includegraphics[width = 1.0\textwidth]{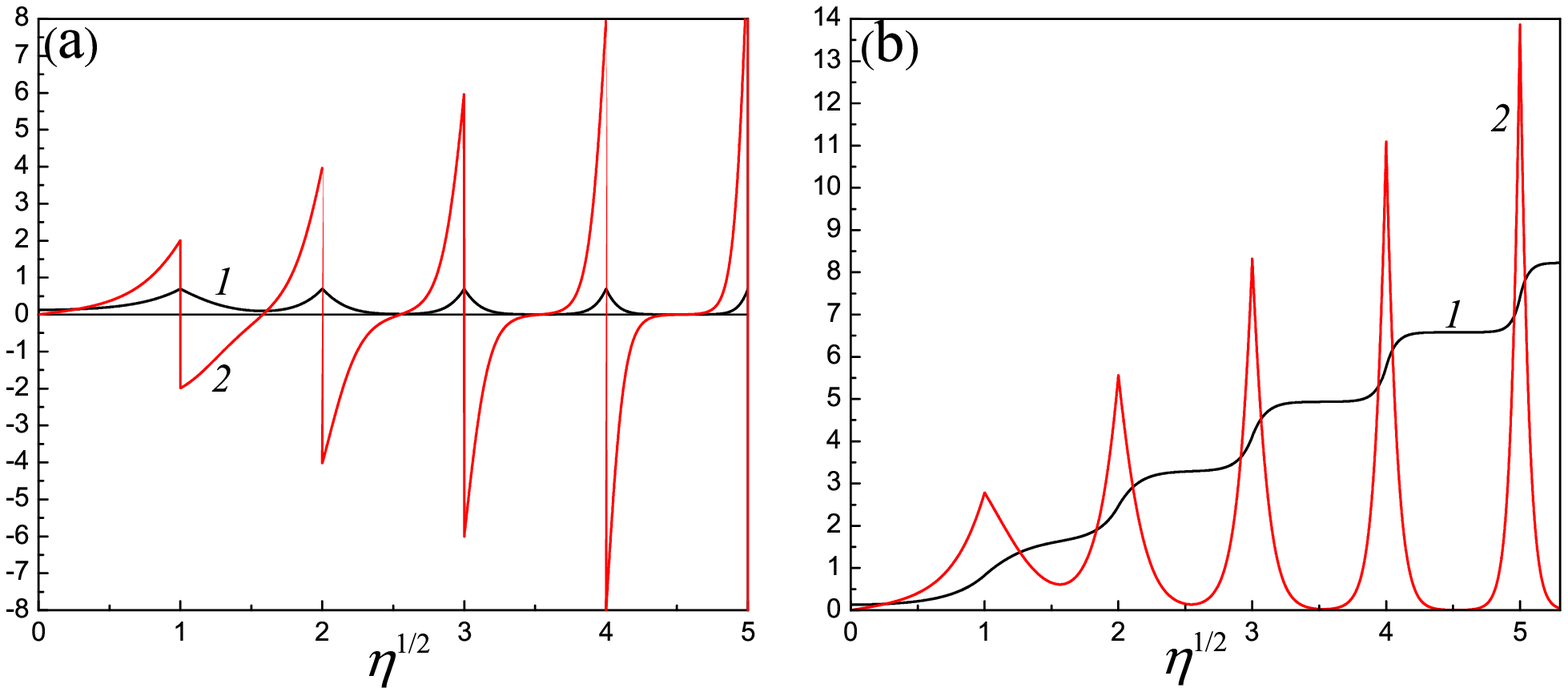} 
\vspace{-6mm}
\caption{\label{fig04} %
Graphs of the functions $\Psi_1''(\tau,\eta)$, $\Psi_2''(\tau,\eta)$ and their derivatives at $\tau=0.5$. \newline %
(a) The functions $\Psi_1''(\tau,\eta)$  ({\it 1}) and $d\Psi_1''(\tau,\eta)\big/d\!\sqrt{\eta}$ \,({\it 2}); %
(b) The functions $\Psi_2''(\tau,\eta)$  ({\it 1}) and $d\Psi_2''(\tau,\eta)\big/d\!\sqrt{\eta}$ \,({\it 2}). %
}%
\end{figure*}

\vspace{0mm} %
\begin{figure*}[t!]
\vspace{0mm} \hspace{-0mm}
\includegraphics[width = 1.0\textwidth]{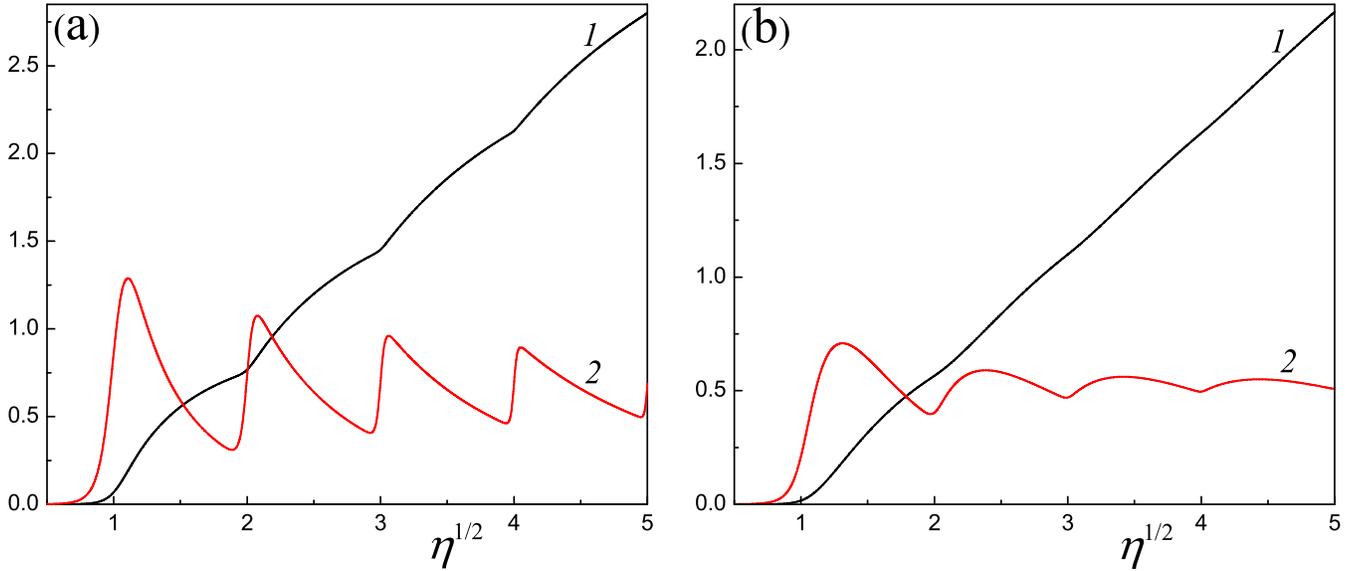} 
\vspace{-6mm}
\caption{\label{fig05} %
Graphs of the functions $(\tau/\eta)\Psi_1(\tau,\eta)$, $2(\tau/\eta)^2\Psi_2(\tau,\eta)$ and their derivatives at $\tau=0.1$. \newline %
(a) The functions $(\tau/\eta)\Psi_1(\tau,\eta)$  ({\it 1}) and $d[(\tau/\eta)\Psi_1(\tau,\eta)]\big/d\!\sqrt{\eta}$ \,({\it 2}); %
(b) The functions $2(\tau/\eta)^2\Psi_2(\tau,\eta)$  ({\it 1}) and $d\big[2(\tau/\eta)^2\Psi_2(\tau,\eta)\big]\big/d\!\sqrt{\eta}$ \,({\it 2}). %
}%
\end{figure*}

More interesting is the case $\eta> 1$, which is realized at low
temperatures. Then the considered functions can be represented in
the form
\begin{equation} \label{41}
\begin{array}{ll}
\displaystyle{%
   \Psi_1(\tau,\eta)=\frac{\eta}{\tau}\Psi_1'(\eta)+\Psi_1''(\tau,\eta), %
}
\end{array}
\end{equation}
\begin{equation} \label{42}
\begin{array}{ll}
\displaystyle{%
   \Psi_2(\tau,\eta)=\frac{\eta^2}{2\tau^2}\Psi_2'(\eta)+\Psi_2''(\tau,\eta). %
}
\end{array}
\end{equation}
Here the functions
\begin{equation} \label{43}
\begin{array}{ll}
\displaystyle{%
   \Psi_1'(\eta)=[x_0]\left\{1-\frac{1}{6\eta}\big([x_0]+1\big)\big(2[x_0]+1\big)\right\}, %
}
\end{array}
\end{equation}
\begin{equation} \label{44}
\begin{array}{ll}
\hspace{-2mm}
\displaystyle{%
  \Psi_2'(\eta)=[x_0] %
  \bigg\{ 1-\frac{1}{3\eta}\big([x_0]+1\big)\big(2[x_0]+1\big)\, +  %
}\vspace{2mm}\\ %
\displaystyle{\hspace{0mm}%
   +\frac{1}{30\eta^2}\big([x_0]+1\big)\big(2[x_0]+1\big)\big(3[x_0]^2+3[x_0]-1\big)\bigg\}%
}
\end{array}
\end{equation}
determine the state of the system at zero temperature. For brevity
here and in the following it is used the designation
$x_0\equiv\sqrt{\eta}$, $x_0>1$ at that, and $[x_0]$ designates the
whole part of a number $x_0$. Graphs of the functions (\ref{43}),
(\ref{44}) and their derivatives are shown in Fig.\,3a and 3b. The
function $\Psi_1'(\eta)$ is continuous and monotonically increasing,
and its derivative at the specific points $x_0=[x_0]$ undergoes jumps %
$\Delta\big(d\Psi_1'(\eta)/d\eta\big)\!\big|_{[x_0]}=1\big/[x_0]^2$
(Fig.\,3a). At $\eta\gg 1$ it has the asymptote
$\Psi_1'(\eta)\approx (2/3)\sqrt{\eta}$. The function
$\Psi_2'(\eta)$ is also continuous and monotonically increasing,
with the asymptote $\Psi_2'(\eta)\approx (8/15)\sqrt{\eta}$ \,at
$\eta\gg 1$. The derivative of the function $\Psi_2'(\eta)$ is an
oscillating function with a varying amplitude (Fig.\,3b).

The functions
\begin{equation} \label{45}
\begin{array}{ll}
\hspace{-0mm}
\displaystyle{%
  \Psi_1''(\tau,\eta)=\Phi_1\!\Big[\tau^{-1}\!\big([x_0]^2-\eta\big)\Big] + %
}\vspace{3mm}\\ %
\displaystyle{\hspace{13mm}%
   +\,\Phi_1\!\Big[\tau^{-1}\!\big(\eta-([x_0]+1)^2\big)\Big] + \bar{\Psi}_1^{\rm{exp}}(\tau,\eta),%
}
\end{array}
\end{equation}
\begin{equation} \label{46}
\begin{array}{ll}
\hspace{-0mm}
\displaystyle{%
  \Psi_2''(\tau,\eta)=\frac{\pi^2}{6}[x_0]-\Phi_2\!\Big[\tau^{-1}\!\big([x_0]^2-\eta\big)\Big] + %
}\vspace{3mm}\\ %
\displaystyle{\hspace{13mm}%
   +\,\Phi_2\!\Big[\tau^{-1}\!\big(\eta-([x_0]+1)^2\big)\Big]\! + \bar{\Psi}_2^{\rm{exp}}(\tau,\eta)%
}
\end{array}
\end{equation}
describe the temperature dependencies of thermodynamic quantities at
low temperatures. The form of the exponentially small at $\tau\ll 1$
functions $\bar{\Psi}_s^{\rm{exp}}(\tau,\eta)$ is given in
Appendix\,A (the formula (\ref{A11})). The dependencies of the
functions (\ref{45}), (\ref{46}) and their derivatives on the
chemical potential are shown in Fig.\,4a and 4b. The function
$\Psi_1''(\tau,\eta)$ has an oscillating character and its
derivative undergoes jumps at $x_0=[x_0]$ (Fig.\,4a). The function
$\Psi_2''(\tau,\eta)$ is monotonically increasing and its derivative
has oscillations (Fig.\,4b). The dependencies of the total functions 
(\ref{41}), (\ref{42}) and their derivatives on the chemical
potential are shown in Fig.\,5a and 5b.

Although, as it was noted, the function  $\Psi_1''(\tau,\eta)$ has
an oscillation form, the total function $\Psi_1(\tau,\eta)$ proves
to be monotonically increasing (Fig.\,5a). Also monotonically
increasing is the function $\Psi_2(\tau,\eta)$ (Fig.\,5b). The
derivatives of both of these functions have an oscillating character
at not large values of the chemical potential.

Thus, the functions $\Psi_1(\tau,\eta)$, $\Psi_2(\tau,\eta)$
themselves through which the thermodynamic quantities are expressed
are not oscillating, in particular there are absent oscillations of
the thermodynamic potential (\ref{11}) on the chemical potential
$\eta$. However, as we will see, dependencies of some quantities on
the chemical potential that include the derivatives of these
functions, such as for example compressibilities, can have a
nonmonotonic character. %

$\vspace{0mm}$

\section{Thermodynamic quantities \newline at low temperatures } %
The most interesting region where quantum effects can manifest
themselves on the macroscopic level is the region of low
temperatures. Let us consider the behavior of the observable
characteristics at low temperatures, such that $\tau\ll 1$. In this
limit, with account of the main exponential corrections
\begin{equation} \label{47}
\begin{array}{ll}
\hspace{-0mm}
\displaystyle{%
  \Psi_1(\tau,\eta)=\frac{\eta}{\tau}\Psi_1'(\eta)+\exp\!\Big[\tau^{-1}\!\big([x_0]^2-\eta\big)\Big] + %
}\vspace{3mm}\\ %
\displaystyle{\hspace{29mm}%
   +\,\exp\!\Big[\tau^{-1}\!\big(\eta-([x_0]+1)^2\big)\Big],%
}
\end{array}
\end{equation}
\begin{equation} \label{48}
\begin{array}{ll}
\hspace{-0mm}
\displaystyle{%
  \Psi_2(\tau,\eta)=\frac{\eta^2}{2\tau^2}\Psi_2'(\eta) + \frac{\pi^2}{6}[x_0] -\exp\!\Big[\tau^{-1}\!\big([x_0]^2-\eta\big)\Big] + %
}\vspace{3mm}\\ %
\displaystyle{\hspace{35mm}%
   +\,\exp\!\Big[\tau^{-1}\!\big(\eta-([x_0]+1)^2\big)\Big].%
}
\end{array}
\end{equation}
For the reduced entropy, we obtain in this approximation:
\begin{equation} \label{49}
\begin{array}{ll}
\hspace{-0mm}
\displaystyle{%
  \tilde{S}= \frac{\pi^2}{3}[x_0]\,\tau - (2\tau-a_1)\,e^{a_1/\tau} + (2\tau-a_2)\,e^{a_2/\tau}. %
}
\end{array}
\end{equation}
Here and below for brevity the designations are used %
$a_1\equiv [x_0]^2-\eta$, $a_2\equiv \eta - ([x_0]+1)^2$. %
The dependencies of the entropy on temperature and density are shown
in Fig.\,6. The dependence on temperature proves to be linear as in
the volume case, but its slope changes by jumps as the energetic
levels are being filled up (Fig.\,6a). With varying the chemical
potential or the density at low temperatures the entropy undergoes
jumps, which are becoming more indistinct as temperature increases
and entirely disappear at rather high temperatures (Fig.\,6b). The
correction to the linear law, in contrast to the volume case where
it is proportional to $T^3$, in this case is exponentially small.

The value of the entropy jump per unit of area depends only on
temperature and is determined by the formula
\begin{equation} \label{50}
\begin{array}{ll}
\displaystyle{%
  \frac{\Delta S}{A} = \frac{\pi m}{3\hbar^2}\,T. %
}
\end{array}
\end{equation}
Pay attention that this quantity does not explicitly depend on the
distance $L$, though certainly this parameter enters into the
condition of applicability of the formula (\ref{50}) %
$T\ll \pi^2\hbar^2\big/2mL^2$. The entropy jumps are accompanied by
the absorption of heat $\Delta Q=T\Delta S$. For electrons the heat
absorbed at the jump of the entropy per unit
of area $\Delta Q/A\approx 1.6\!\cdot\!10^{-5}\,\,T^2$ %
${\rm erg}/\!({\rm cm^2K^2})$ (temperature in Kelvins). For $^3$He
atoms this quantity is by three orders greater.

\vspace{0mm} %
\begin{figure*}[t!]
\vspace{0mm} \hspace{-0mm}
\includegraphics[width = 1.0\textwidth]{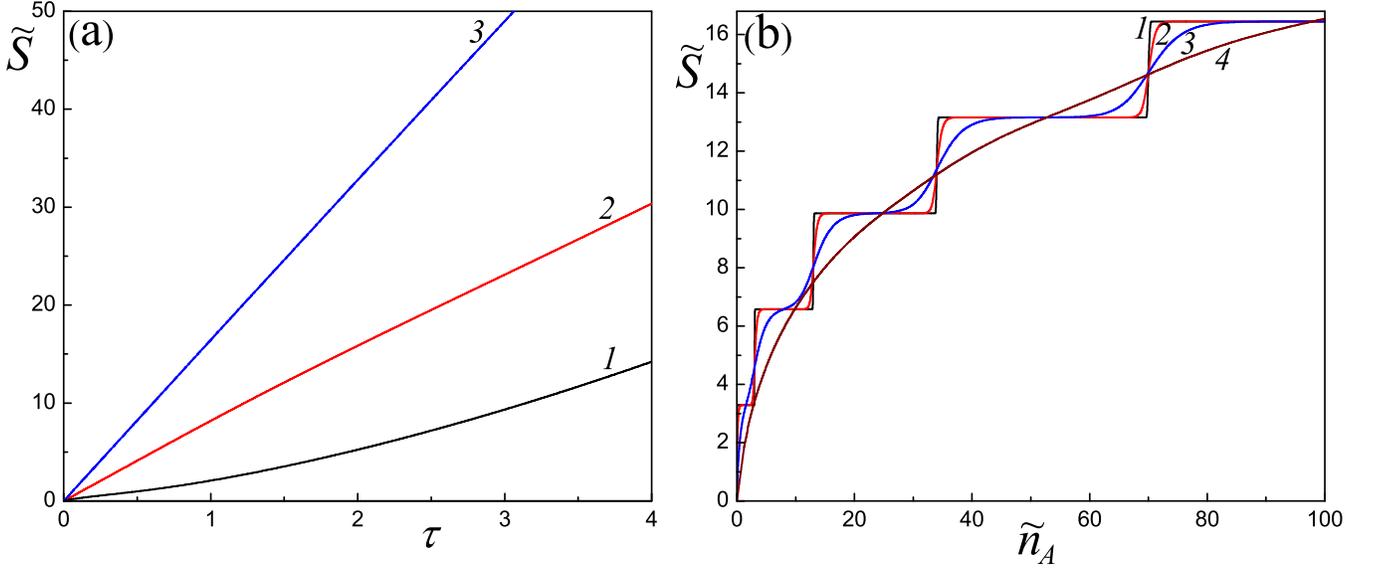} 
\vspace{-8mm}
\caption{\label{fig06} %
(a) The temperature dependencies of the reduced entropy $\tilde{S}(\tau;x_0)$ %
at a fixed value of the chemical potential $x_0\equiv\sqrt{\eta}$: %
({\it 1}) $x_0=1.1$;  ({\it 2}) $x_0=3.0$;  ({\it 3}) $x_0=5.5$;  %
(b) The dependencies of the quantity $\tilde{S}(\tilde{n}_{\!A};\tau)/\tau$ %
on the reduced density at fixed temperature: %
({\it 1}) $\tau=0.01$; ({\it 2}) $\tau=0.1$; ({\it 3}) $\tau=0.5$; ({\it 4}) $\tau=2.0$. %
}%
\end{figure*}

The density at low temperatures has only the exponentially small,
depending on temperature, correction
\begin{equation} \label{51}
\begin{array}{ll}
\displaystyle{%
  \tilde{n}=\tilde{n}_{\!A} =  \eta\Psi_1'(\eta)+\tau\left(e^{a_1/\tau}+e^{a_2/\tau}\right). %
}
\end{array}
\end{equation}
Neglecting the exponential corrections, the chemical potential in
the expression for the entropy (\ref{49}) and in other thermodynamic
quantities can be taken with a good accuracy at zero temperature. In
this approximation the chemical potential is connected with the
reduced densities by the relation
\begin{equation} \label{52}
\begin{array}{ll}
\displaystyle{%
  \tilde{n}=\tilde{n}_{\!A} =  [x_0]\left\{\eta-\frac{1}{6}\big([x_0]+1\big)\big(2[x_0]+1\big)\right\}. %
}
\end{array}
\end{equation}\newpage\noindent
The density monotonically increases with increasing the chemical
potential, undergoing breaks (discontinuities in the derivative)
at the specific points $x_0=[x_0]$. %
In the limit of rather high density when $x_0\gg 1$, we can set
$[x_0]\approx x_0$ and from (\ref{52}) there follows the usual
formula that relates the density of the bulk Fermi gas with the
chemical potential at zero temperature: $n=(2m\mu)^{3/2}\!\big/3\pi^2\hbar^3$. %
The thermodynamical potential in this limit also acquires the usual form %
$\Omega=-AL\frac{4\sqrt{2}}{15\pi^2}\,m^{3/2}\mu^{5/2}\!\big/\hbar^3$.
The condition $x_0\gg 1$ is equivalent to the condition $N^{1/3}\gg (A/L^2)^{1/3}$. %
In the volume case $A\approx L^2$ it is equivalent to the condition
$N^{1/3}\gg 1$, which is always true in a system of large number of
particles. The exact formulas should be used under fulfilment of the
condition $N \sim A/L^2$.

\vspace{-1mm} %
If thermodynamic quantities are taken in the reduced form
(\ref{24}), and the dimensionless chemical potential $\eta$ and the
dimensionless temperature $\tau$ are used as independent variables,
then as it was shown the geometrical dimensions fall out of the
thermodynamic relations, in particular the distance between plates
$L$ falls out (or the thickness of a film, from experimentalist's
point of view). Meanwhile, exactly the dependencies of the
observable quantities on the thickness of a film are of interest in
experiment. To obtain such dependencies, the relations derived above
should be presented in the dimensional form. At that, the form of
dependence of the thermodynamic quantities on the thickness of a
film will essentially depend on what quantity is being fixed when
studying such dependencies: the total density $n$ or the surface
density $n_{\!A}$. Let us show it on the example of dependence of
the chemical potential on the thickness of a film.

$\vspace{8mm}$

The formula (\ref{52}) can be written in the form
\begin{equation} \label{53}
\begin{array}{ll}
\displaystyle{%
  \frac{2n}{\pi}L^3=\frac{2n_{\!A}}{\pi}L^2 =  [x_0]\left\{x_0^2-\frac{1}{6}\big([x_0]+1\big)\big(2[x_0]+1\big)\right\}. %
}
\end{array}
\end{equation}
As we see, the dependence of the distance $L$ on the parameter $x_0$
will be different depending on what is fixed -- the volume or the
surface density.  The chemical potential is expressed through $L$
and the parameter $x_0$ by the formula
\begin{equation} \label{54}
\begin{array}{ll}
\displaystyle{%
  \mu=\frac{\pi^2\hbar^2}{2mL^2}\,x_0^2. %
}
\end{array}
\end{equation}
The formulas (\ref{53}) and (\ref{54}) define parametrically
(parameter $x_0$) the dependence of the chemical potential on the
thickness of a film at fixed $n$ or $n_{\!A}$. It is easy to check
that at a fixed surface density $(d\mu/dL)_{n_{\!A}}<0$ and,
therefore, the chemical potential monotonically decreases with
increasing $L$. At a fixed volume density the derivative
$(d\mu/dL)_{n}$ turns into zero at the minimum points determined by
the equation $x_0^2-\frac{1}{2}\big([x_0]+1\big)\big(2[x_0]+1\big)=0$. %
The distance between planes at the minimum points of the chemical
potential is determined by the formula
\begin{equation} \label{55}
\begin{array}{ll}
\displaystyle{%
  \frac{2n}{\pi}L_{\rm{min}}^3 = \frac{1}{3}\,[x_0]\big([x_0]+1\big)\big(2[x_0]+1\big). %
}
\end{array}
\end{equation}
At the points $x_0=[x_0]\geq 2$ the sign of the derivative
$(d\mu/dL)_{n}$ changes by a jump. These points correspond to the
local maximums of the chemical potential, so that
\begin{equation} \label{56}
\begin{array}{ll}
\displaystyle{%
  \frac{2n}{\pi}L_{\rm{max}}^3 = \frac{1}{6}\,[x_0]\big([x_0]-1\big)\big(4[x_0]+1\big). %
}
\end{array}
\end{equation}\newpage\noindent
Thus, the dependence of the chemical potential on $L$ at a fixed
volume density has an oscillating character. The dependencies
$\mu=\mu(L)$ at fixed surface (curve {\it 1}) and volume (curve {\it
2}) densities are shown in Fig.\,7.

\vspace{0mm} %
\begin{figure}[t!]
\vspace{0mm} \hspace{-0mm}
\includegraphics[width = 1.0\columnwidth]{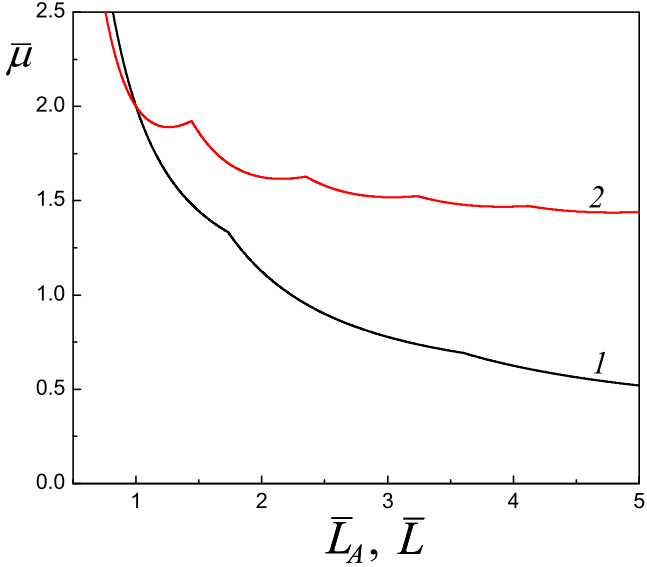} 
\vspace{-8mm}
\caption{\label{fig07} %
The dependencies of the chemical potential on the distance between planes: %
({\it 1}) $\bar{\mu}=\bar{\mu}(\bar{L}_A)$ at a fixed surface density $n_{\!A}$, here %
$\bar{\mu}\equiv(m/\pi\hbar^2n_{\!A})\mu$, $\bar{L}_A\equiv(2n_{\!A}/\pi)^{1/2}L$; %
({\it 2}) $\bar{\mu}=\bar{\mu}(\bar{L})$ at a fixed volume density $n$, here %
$\bar{\mu}\equiv(2/\pi^4)^{1/3}(m/\hbar^2n^{2/3})\mu$, $\bar{L}\equiv(2n/\pi)^{1/3}L$. %
}%
\end{figure}

\vspace{0mm} %
\begin{figure}[b!]
\vspace{0mm} \hspace{-0mm}
\includegraphics[width = 1.0\columnwidth]{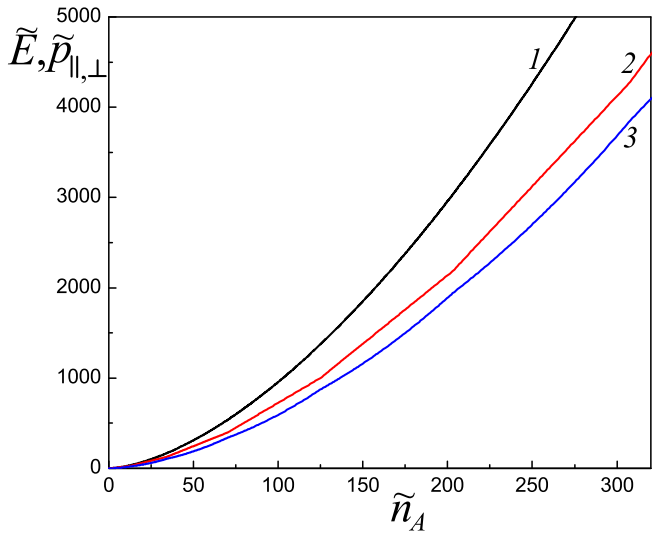} 
\vspace{-8mm}
\caption{\label{fig08} %
The dependencies of the reduced energy and pressures on the reduced
density at zero temperature: \newline %
({\it 1}) $\tilde{E}(\tilde{n}_{\!A})$; %
({\it 2}) $\tilde{p}_\perp(\tilde{n}_{\!A})$; %
({\it 3}) $\tilde{p}_\parallel(\tilde{n}_{\!A})$.
}%
\end{figure}

\begin{figure}[t!]
\vspace{-2mm} \hspace{-0mm}
\includegraphics[width = 1.0\columnwidth]{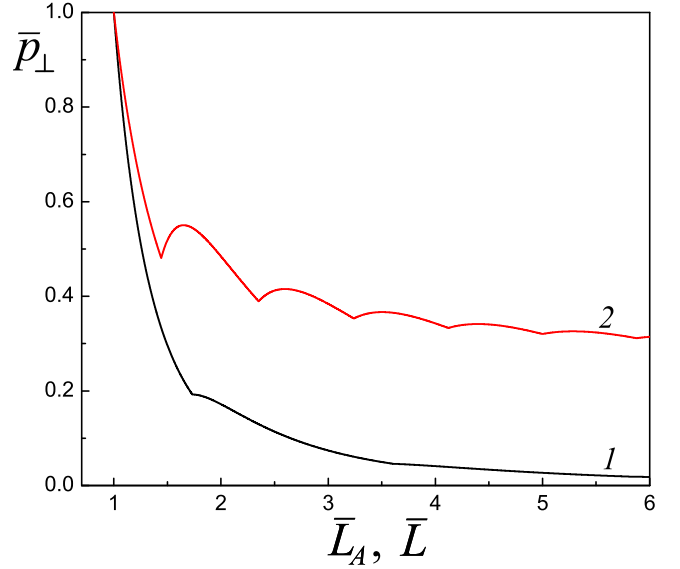} 
\vspace{-8mm}
\caption{\label{fig09} %
The dependencies of the perpendicular pressure on the distance between planes: %
({\it 1}) $\bar{p}_\perp=\bar{p}_\perp(\bar{L}_A)$ at a fixed surface density $n_{\!A}$, here 
$\bar{p}_\perp\equiv(2m/\pi^3\hbar^2)(\pi/2n_{\!A})^{5/2}$, $\bar{L}_A\equiv(2n_{\!A}/\pi)^{1/2}L$; %
({\it 2}) $\bar{p}_\perp=\bar{p}_\perp(\bar{L})$ at a fixed volume density $n$, here 
$\bar{p}_\perp\equiv(2m/\pi^3\hbar^2)(\pi/2n)^{5/3}$, $\bar{L}\equiv(2n/\pi)^{1/3}L$. %
}%
\end{figure}

The pressures at low temperatures are determined by the formulas: 
\begin{equation} \label{57}
\begin{array}{ll}\hspace{-1mm}
\displaystyle{%
  \tilde{p}_\parallel=\frac{\eta^2}{4}\Psi_2'(\eta)+\frac{\pi^2}{12}[x_0]\tau^2 %
  +\frac{\tau^2}{2}\!\left(-e^{a_1/\tau}\!+e^{a_2/\tau}\right), %
}
\end{array}
\end{equation}
\begin{equation} \label{58}
\begin{array}{ll}\hspace{0mm}
\displaystyle{%
  \tilde{p}_\perp=\eta^2\big[\Psi_1'(\eta)-\Psi_2'(\eta)\big] + %
}\vspace{3mm}\\ %
\displaystyle{\hspace{07mm}%
   +\,\tau\!\left([x_0]^2e^{a_1/\tau}\!+([x_0]+1)^2e^{a_2/\tau}\right). %
}
\end{array}
\end{equation}
As seen, the parallel pressure, in addition to the exponential
temperature correction, also contains the power correction
proportional to the square of temperature, and the perpendicular
pressure contains only the exponential temperature correction. The
dependencies of the reduced energy and pressures on the reduced
density at zero temperature are shown in Fig.\,8. Both the energy
and the pressures monotonically increase with increasing the
chemical potential or the density. At the points where the filling
of discrete levels begins the dependence
$\tilde{p}_\perp=\tilde{p}_\perp(\tilde{n}_{\!A})$ undergoes breaks.
In the limit $x_0\gg 1$ at zero temperature the both pressures
$p_\parallel, p_\perp$ prove to be equal to the pressure of the bulk
degenerate Fermi gas.

There are of interest the dependencies of the perpendicular pressure
on the distance between planes, presented in Fig.\,9. The pressure
at fixed $n_{\!A}$ decreases with increasing $L$ (curve {\it 1}). At
one point, corresponding to $x_0=2$, this curve undergoes a break:
the derivative $dp_\perp/dL$ on the left at this point is negative
and on the right it equals to zero. At a fixed $n$ the dependence
$p_\perp=p_\perp(L)$ has an oscillating form (curve {\it 2}). %
The extremum points of this function can be found in the same way as
for the dependence $\mu(L)$ at fixed $n$ and are given by the expressions: %
\begin{equation} \label{58I}
\begin{array}{ll}\hspace{0mm}
\displaystyle{%
  \frac{2n}{\pi}L_{\rm{min}}^3 = \frac{1}{6}\,[x_0]\big([x_0]-1\big)\big(4[x_0]+1\big), %
}\vspace{2mm}\\ %
\displaystyle{\hspace{0mm}%
   \frac{2n}{\pi}L_{\rm{max}}^3 = \frac{1}{12}\,[x_0]\big([x_0]-1\big)\big(8[x_0]+11\big). %
}
\end{array}
\end{equation}
On the basis of the performed analysis, seemingly, a general
conclusion can be made that the oscillating dependencies of
thermodynamic quantities on the width $L$ take place for the case of
the fixed total density and they are absent when the surface density
is fixed.

\vspace{0mm} %
\begin{figure}[b!]
\vspace{0mm} \hspace{-0mm}
\includegraphics[width = 1.0\columnwidth]{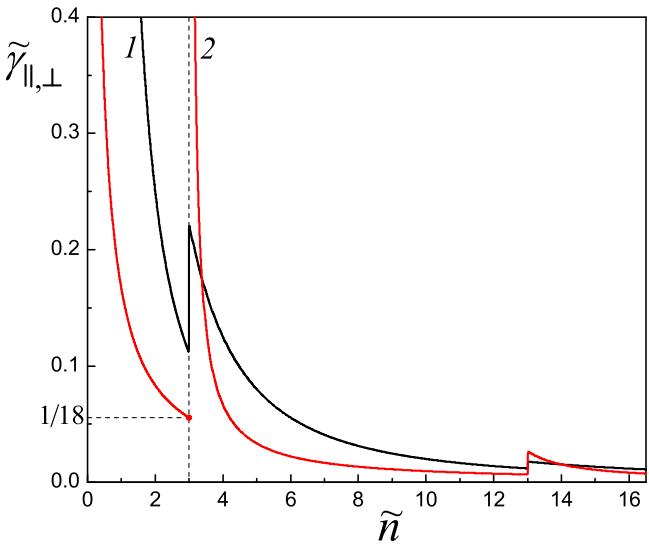} 
\vspace{-8mm}
\caption{\label{fig10} %
The dependencies of compressibilities \newline on the reduced density at zero temperature:  \newline %
({\it 1}) $\tilde{\gamma}_\parallel=\tilde{\gamma}_\parallel(\tilde{n})$; %
({\it 2}) $\tilde{\gamma}_\perp=\tilde{\gamma}_\perp(\tilde{n})$. %
}%
\end{figure}

He we give also the formulas, following from the general relations
(\ref{34})\,--\,(\ref{38}), for the compressibilities at zero
temperature:
\begin{equation} \label{59}
\begin{array}{ll}\hspace{0mm}
\displaystyle{%
  \tilde{\gamma}_\parallel=\frac{[x_0]}{\big\{\eta\Psi_1'(\eta)\big\}^2}= %
}\vspace{2mm}\\ %
\displaystyle{\hspace{05mm}%
   =\frac{1}{\displaystyle{[x_0]\left\{x_0^2-\frac{1}{6}\big([x_0]+1\big)\big(2[x_0]+1\big)\right\}^2}}, %
}
\end{array}
\end{equation}
\begin{equation} \label{60}
\begin{array}{ll}\hspace{0mm}
\displaystyle{%
  \tilde{\gamma}_\perp=\frac{[x_0]}{2\eta^2\Big\{\!\big(3\Psi_1'(\eta)-5\Psi_2'(\eta)\big)[x_0]+2\big(\Psi_1'(\eta)\big)^2 \Big\}}= %
}\vspace{2mm}\\ %
\displaystyle{\hspace{0mm}%
   =\frac{1}{\displaystyle{[x_0]\big([x_0]+1\big)\big(2[x_0]+1\big)\!\left\{\!x_0^2-\frac{7}{9}[x_0]^2-\frac{2}{3}[x_0]+\frac{4}{9}\right\}}}. %
}
\end{array}
\end{equation}
The dependencies of compressibilities on the volume density at zero
temperature are shown in Fig.\,10. As seen, at some values of
density, at which the filling of levels begins, the
compressibilities undergo jumps. On approaching to the point that
corresponds to $x_0=2$ from the side of large densities, the
perpendicular compressibility tends to infinity. The nature of this
divergence is similar to the nature of the break at this point in
the dependence of the perpendicular pressure $p_\perp=p_\perp(L)$
(Fig.\,9, curve {\it 1}). In the limit $x_0\gg 1$ we have %
$\gamma_\parallel=\gamma_\perp=\frac{3^{1/3}}{\pi^{4/3}}\,m\big/\hbar^2n^{5/3}$. %
The quantity $u^2=1\big/mn\gamma_\parallel=\frac{\pi^{4/3}}{3^{1/3}}\,\hbar^2n^{2/3}\!\big/m^2$ %
determines in this case the square of the speed of sound in the bulk Fermi gas. %

In conclusion of this section, we proceed to the analysis of the low
temperature behavior of heat capacities. For the first time, the
calculation of the electron heat capacity for particles of small
size with account of the discreteness of the energy levels was made
by Fr\"{o}hlich \cite{Frohlich}. He showed that, in contrast to the
bulk Fermi gas which has the linear temperature dependence of the
heat capacity, in the case of small in all coordinates particles the
heat capacity decreases exponentially with decreasing temperature.
In the case considered in the present paper, in addition to the
discrete levels there is possible a free motion of particles along
planes, that leads to the maintenance of the linear law in the
temperature dependence in given conditions.

In the main approximation all introduced above heat capacities
(\ref{29})\,--\,(\ref{32}), as it had to be expected, prove to be
identical and proportional to temperature
\begin{equation} \label{61}
\begin{array}{ll}\hspace{0mm}
\displaystyle{%
  \tilde{C} \approx \frac{\pi^2}{3}[x_0]\,\tau. %
}
\end{array}
\end{equation}
Under fulfilment of the condition $x_0\gg 1$ the formula (\ref{61}),
naturally, turns into the expression for the low temperature heat
capacity of the bulk Fermi gas $C=(\pi/3)^{2/3}(mT/\hbar^2)\,n^{1/3}LA$. %
The heat capacity as a function of the chemical potential and
density undergoes jumps at the points, in which the filling of new
discrete levels begins (Fig.\,11). The account for the corrections
to the formula (\ref{61}) smoothes out the steps. The value of the
jump of the heat capacity is the same as that of the entropy
(\ref{50}): $\Delta C/A=\pi mT/3\hbar^2$.

\begin{figure}[b!]
\vspace{0mm} \hspace{-0mm}
\includegraphics[width = 1.01\columnwidth]{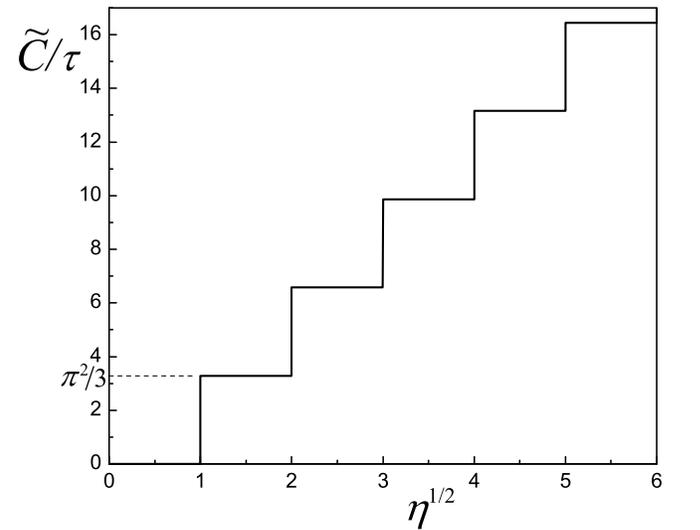} 
\vspace{-7mm}
\caption{\label{fig11} %
The dependence of the ratio of the reduced heat capacity to the dimensionless temperature %
$\tilde{C}\big/\tau = \frac{\pi^2}{3}[x_0]$ at $\tau\ll 1$ on the
chemical potential $\eta^{1/2}$.
}%
\end{figure}

For the heat capacities $\tilde{C}_{AL}$, $\tilde{C}_{A
p_\parallel}$, $\tilde{C}_{L p_\perp}$, $\tilde{C}_{A p_\perp}$ the
corrections to the linear law (\ref{61}) have the exponential
character, and for two heat capacities these corrections are
proportional to $\tau^3$: 
\begin{equation} \label{62}
\begin{array}{ll}\hspace{-2mm}
\displaystyle{%
  \tilde{C}_{p_\parallel p_\perp} \approx \frac{\pi^2}{3}[x_0]\,\tau\,\times %
}\vspace{2mm}\\ %
\displaystyle{\hspace{0mm}%
   \times\!\left\{\!1+\frac{2\pi^2}{15}\frac{[x_0]}
   {\eta^2}\frac{\big(5[x_0]\Psi_2'-3[x_0]\Psi_1'-2\Psi_1'^2\big)}{\Psi_1'\big([x_0]\Psi_2'+\Psi_1'\Psi_2'-2\Psi_1'^2\big)}\,\tau^2\!\right\}, %
}
\end{array}
\end{equation}
\begin{equation} \label{63}
\begin{array}{ll}\hspace{0mm}
\displaystyle{%
  \tilde{C}_{L p_\parallel} \approx \frac{\pi^2}{3}[x_0]\,\tau  %
  \left\{1+\frac{\pi^2}{3}[x_0]\frac{\tau^2}{\big(\eta\Psi_1'\big)^2}\right\}.
}
\end{array}
\end{equation}
The differences between the heat capacities $\tilde{C}_{p_\parallel
p_\perp}$, $\tilde{C}_{L p_\parallel}$ and other heat capacities are
proportional to $\tau^3$.

\section{ Thermodynamic quantities \newline at high temperatures } %
Let us consider the area of high temperatures, where the de Broglie
wavelength is much less than the average distance between particles:
$\Lambda/l=\Lambda n^{1/3}\ll 1$. This condition is fulfilled if the
parameter $t=\eta/\tau$ is negative and large by absolute value
$|t|\gg 1$. At high temperatures and macroscopic distances between
planes, such that $\Lambda/L=2/\sqrt{\pi\tau}\ll 1$, also the
condition $\sqrt{\tau}\gg 1$ holds. In this approximation, in the
sum of the formula (\ref{40}) it is sufficient to account for the
main term with $l=1$ . Taking into account the relation
$\theta_3(0,q)=\big[2K(m)/\pi\big]^{1/2}$, we obtain
\begin{equation} \label{64}
\begin{array}{ll}\hspace{0mm}
\displaystyle{%
  \Psi_s(\tau,\eta)= \frac{1}{2}  %
  \left[\sqrt{\frac{2K(m)}{\pi}}-1\right]e^{\eta/\tau}.
}
\end{array}
\end{equation}
Here $K(m)$ is the full elliptic integral of the first kind, and
$q\equiv q(m)=\exp{\!\big[\!-\!\pi K(1-m)\big/K(m)\big]}$ is the
Jacobi parameter \cite{AS}. Considering the definition of the
theta-function in (\ref{40}), we find that the parameter $m$ and the
dimensionless temperature are connected by the relation
\begin{equation} \label{65}
\begin{array}{ll}\hspace{0mm}
\displaystyle{%
  \frac{1}{\tau}=\pi\frac{K(1-m)}{K(m)}.  %
}
\end{array}
\end{equation}
For considered large values of $\tau$, the parameter $m$ is close to
unity. Using expansions in the small parameter $m_1=1-m$ and taking
into account that $K(m)\approx\frac{1}{2}\ln(16/m_1)$, from
(\ref{65}) we have: $m_1=16e^{-\pi^2\tau}$. For $\sqrt{\tau}\gg 1$,
taking account of the main terms, we obtain
\begin{equation} \label{66}
\begin{array}{ll}\hspace{0mm}
\displaystyle{%
  \Psi_s(\tau,\eta)= \frac{1}{2}  %
  \big[\sqrt{\pi\tau}-1\big]e^{\eta/\tau}.
}
\end{array}
\end{equation}
Taking account of only the first term in square brackets in
(\ref{66}) leads to the relations for the classical ideal gas, and
accounting for the second term in brackets gives the correction on
the finite width $L$ proportional to the ratio $\Lambda/L$. The
thermodynamic potential in the classical limit with account of such
correction acquires the form
\begin{equation} \label{67}
\begin{array}{ll}
\displaystyle{\hspace{0mm}%
  \Omega=-\frac{2TAL}{\Lambda^3}\left(1-\frac{\Lambda}{2L}\right)e^{\mu/T}. %
}
\end{array}
\end{equation}
From the formula for the number of particles
\begin{equation} \label{68}
\begin{array}{ll}
\displaystyle{\hspace{0mm}%
  N=-\!\left( \frac{\partial \Omega}{\partial \mu}\right)_{\!T,A,L}= %
  \frac{2AL}{\Lambda^3}\left(1-\frac{\Lambda}{2L}\right)e^{\mu/T} %
}
\end{array}
\end{equation}
there follows the dependence of the chemical potential on the
density and temperature:
\begin{equation} \label{69}
\begin{array}{ll}
\displaystyle{\hspace{0mm}%
  \frac{\mu}{T}=\ln\!\left( \frac{n\Lambda^3}{2}\right) + \frac{\Lambda}{2L}.  %
}
\end{array}
\end{equation}
For the entropy we have the expression that generalizes the
Sackur-Tetrode formula \cite{LL} to account for the quantum size effect: %
\begin{equation} \label{70}
\begin{array}{ll}
\displaystyle{\hspace{0mm}%
  S=N\ln\!\frac{2e^{5/2}}{n\Lambda^3} - N\frac{\Lambda}{4L}.  %
}
\end{array}
\end{equation}
In the parallel pressure the linear with respect to $\Lambda/L$
correction is absent, and it is present in the perpendicular pressure: %
\begin{equation} \label{71}
\begin{array}{ll}
\displaystyle{\hspace{0mm}%
  p_\parallel=nT, \qquad
  p_\perp=nT\!\left(1+\frac{\Lambda}{2L}\right). %
}
\end{array}
\end{equation}
Here is also the formula for the energy with account of such correction: %
\begin{equation} \label{72}
\begin{array}{ll}
\displaystyle{\hspace{0mm}%
  E=\frac{3}{2}\,nT\!\left(1+\frac{\Lambda}{6L}\right). %
}
\end{array}
\end{equation}
Let us write down in the high temperature limit the expressions for
all heat capacities, which were defined above, with account of the
quantum size correction:
\begin{equation} \label{73}
\begin{array}{ll}
\displaystyle{\hspace{0mm}%
  C_{AL}=N\!\left(\frac{3}{2}+\frac{\Lambda}{8L}\right),\quad %
  C_{p_\parallel p_\perp}=\frac{5}{2}\,N,
}\vspace{2mm}\\ %
\displaystyle{\hspace{0mm}%
   C_{Lp_\parallel}=N\!\left(\frac{5}{2}+\frac{\Lambda}{8L}\right),\quad %
   C_{Ap_\parallel}=N\!\left(\frac{5}{2}+\frac{3\Lambda}{8L}\right), %
}\vspace{2mm}\\ %
\displaystyle{\hspace{0mm}%
   C_{Lp_\perp}=N\!\left(\frac{5}{2}-\frac{\Lambda}{8L}\right),\quad %
   C_{Ap_\perp}=N\!\left(\frac{5}{2}-\frac{3\Lambda}{8L}\right). %
}
\end{array}
\end{equation}
As seen, in the volume limit $C_{AL}$ turns into the heat capacity
at a constant volume of the classical ideal gas $C_V=(3/2)N$, and
the other five heat capacities turn into the heat capacity of the
classical gas at a constant pressure $C_p=(5/2)N$. It should be also
noted that the heat capacity $C_{p_\parallel p_\perp}$ does not
contain the linear in the parameter $\Lambda/L$ correction.

\section{Conclusion}\vspace{0mm}
In the paper there have been derived the exact formulas for
calculation of the thermodynamic functions of the ideal Fermi gas in
the quantum well formed by two parallel walls. It is shown that all
thermodynamic quantities, written in the dimensionless reduced form
not containing the geometric dimensions, can be expressed through
some standard functions of the dimensionless temperature and the
dimensionless chemical potential and their derivatives. These
functions themselves do not oscillate with varying the chemical
potential or density, but the derivatives of these functions have
oscillations (Fig.\,5). Through the introduced standard functions
there are calculated the thermodynamic potential, energy, density,
entropy, equations of state, heat capacities and compressibilities
of the Fermi gas at arbitrary temperatures in the considered
conditions of the confined geometry. It is shown that owing to the
anisotropy the Fermi gas in this case has two equations of state
since the pressures perpendicular and parallel to planes are
different, and also is characterized by a set of several  heat
capacities. At low temperatures the entropy and all heat capacities
depend on temperature in the same linear way and undergo jumps at
the beginning of the filling of new discrete energy levels. It is
shown that the character of dependence of thermodynamic quantities
on the distance between planes essentially depends on whether this
dependence is considered at a fixed surface or at a fixed volume
density. At a fixed surface density the thermodynamic quantities
vary monotonically with the distance between planes, and at a fixed
volume density they undergo oscillations. In the area of high
temperatures the quantum corrections to thermodynamic quantities are
obtained, which are proportional to the ratio of the thermal de
Broglie wavelength to the distance between planes.

\appendix

\section{Calculation of functions $\Psi_s(\tau,\eta)$}\label{A}
\vspace{-3mm}

First, let us consider the case when $\eta<1$. Using the formula at
$t<0$ that is correct in this case
\begin{equation} \label{A01}
\begin{array}{ll}
\displaystyle{\hspace{0mm}%
  \Phi_s(t)=\sum_{l=1}^\infty (-1)^{l+1}\frac{e^{lt}}{l^s}, %
}
\end{array}
\end{equation}
we obtain
\begin{equation} \label{A02}
\begin{array}{ll}
\displaystyle{%
   \Psi_s(\tau,\eta)=\frac{1}{2}\sum_{l=1}^\infty\!
   \frac{(-1)^{l+1}}{l^s}\,e^{l\eta/\tau}\theta_3\!\!\left(0,\frac{l}{\pi^2\tau}\right)\! %
   -\frac{1}{2}\,\Phi_s\left(\frac{\eta}{\tau}\right), %
}
\end{array}
\end{equation}
where $\theta_3(\nu,x)\equiv 1+2\sum_{k=1}^\infty  e^{-k^2\pi^2 x}\cos 2\pi\nu x$\, %
is the theta-function, so that
\begin{equation} \label{A03}
\begin{array}{ll}
\displaystyle{\hspace{0mm}%
  \theta_3\!\!\left(0,\frac{l}{\pi^2\tau}\right)\equiv 1+2\sum_{n=1}^\infty e^{-ln^2/\tau}. %
}
\end{array}
\end{equation}
Note the useful relations:
\begin{equation} \label{A04}
\begin{array}{ll}
\displaystyle{\hspace{0mm}%
  \Phi_1(t)=\ln\!\big(1+e^t\big), \quad  %
  \frac{d\Phi_1(t)}{dt}=\frac{e^t}{1+e^t},
}\vspace{2mm}\\ %
\displaystyle{\hspace{0mm}%
   \Phi_2(t)=\frac{t^2}{2}\!\left(1+\frac{\pi^2}{3t^2}\right)\! - \Phi_2(-t). %
}
\end{array}
\end{equation}

Somewhat more complex is the case $\eta>1$, realized at low
temperatures. We consider that for $t>0$ and $s\geq 1$: 
\begin{equation} \label{A05}
\begin{array}{ll}
\displaystyle{\hspace{0mm}%
  \Phi_s(t)= \frac{t^s}{\Gamma(s+1)}\big[1+\chi_s(t)\big]+(-1)^{s-1}\Phi_s(-t),  %
}
\end{array}
\end{equation}
where
\begin{equation} \label{A06}
\begin{array}{ll}
\displaystyle{\hspace{-1mm}%
  \chi_s(t)\equiv s(s-1)!\sum_{l=1}^{s-1}\!\frac{\big[(-1)^l\!-\!1\big]\big(2^{-l}\!-\!1\big)\zeta(l+1)}  %
  {(s\!-\!1\!-\!l)!\,t^{l+1}},%
}
\end{array}
\end{equation}
$\Gamma(s)$ is the gamma function, $\zeta(l)$ is the Riemann zeta
function.

The functions (\ref{12}) can be written in the form
\begin{equation} \label{A07}
\begin{array}{ll}
\displaystyle{%
  \Psi_s(\tau,\eta)\equiv\sum_{n=1}^\infty \Phi_s\Big[\tau^{-1}\big(\eta-n^2\big)\Big]= %
}\vspace{2mm}\\ %
\displaystyle{\hspace{0mm}%
   =\sum_{n=1}^{[x_0]} \Phi_s\Big[\tau^{-1}\!\big(\eta-n^2\big)\Big] + %
    \sum_{n=[x_0]+1}^\infty\!\Phi_s\Big[\tau^{-1}\!\big(\eta-n^2\big)\Big], %
}
\end{array}
\end{equation}
where the designation is used $x_0\equiv\sqrt{\eta}=\!\sqrt{\mu/\varepsilon_L}$. %

In the first term the argument $\tau^{-1}\big(\eta-n^2\big)>0$ and
the formula (\ref{A05}) can be used, and in the second term
$\tau^{-1}\big(\eta-n^2\big)<0$ and the expansion (\ref{A01}) is
valid, so that
\begin{equation} \label{A08}
\begin{array}{ll}
\displaystyle{\hspace{0mm}%
  \Psi_s(\tau,\eta)= %
}\vspace{1mm}\\ %
\displaystyle{\hspace{3mm}%
  =\!\frac{1}{\Gamma(s+1)}\sum_{n=1}^{[x_0]}\tau^{-s}\big(\eta-n^2\big)^{\!s}  %
  \Big\{\!1+\chi_s\Big[\tau^{-1}\!\big(\eta-n^2\big)\Big]\!\Big\}+ %
}\vspace{2mm}\\ %
\displaystyle{\hspace{57mm}%
  +\,\Psi_s^{\rm{exp}}(\tau,\eta),%
}
\end{array}
\end{equation}
where the function
\begin{equation} \label{A09}
\begin{array}{ll}
\displaystyle{\hspace{0mm}%
  \Psi_s^{\rm{exp}}(\tau,\eta)\equiv %
  (-1)^{s+1}\sum_{n=1}^{[x_0]} \Phi_s\Big[\tau^{-1}\!\big(n^2-\eta\big)\Big]\,+ %
}\vspace{1mm}\\ %
\displaystyle{\hspace{22mm}%
  + \sum_{n=[x_0]+1}^\infty\Phi_s\Big[\tau^{-1}\!\big(\eta-n^2\big)\Big] %
}
\end{array}
\end{equation}
contains the exponential terms, which are small for $\tau\ll 1$ but,
however, they can be considerable near the specific points at $\eta=n^2$. %
In (\ref{A09}) the main contribution is given by the terms with $n=[x_0]$ and $n=[x_0]+1$, %
and other terms give the exponentially small contribution of higher
order. Therefore, after having extracted the main contribution, the
formula (\ref{A09}) can be written in the form
\begin{equation} \label{A10}
\begin{array}{ll}
\displaystyle{\hspace{0mm}%
  \Psi_s^{\rm{exp}}(\tau,\eta)= %
  (-1)^{s+1}\Phi_s\Big[\tau^{-1}\!\big([x_0]^2-\eta\big)\Big]\,+ %
}\vspace{2mm}\\ %
\displaystyle{\hspace{11mm}%
  + \,\Phi_s\Big[\tau^{-1}\!\big(\eta-\!([x_0]+1)^2\big)\Big]\! +  %
  \bar{\Psi}_s^{\rm{exp}}(\tau,\eta),
}
\end{array}
\end{equation}
where       
\begin{equation} \label{A11}
\begin{array}{ll}
\displaystyle{\hspace{0mm}%
  \bar{\Psi}_s^{\rm{exp}}(\tau,\eta)\equiv %
  (-1)^{s+1}\sum_{n=1}^{[x_0]-1} \Phi_s\Big[\tau^{-1}\!\big(n^2-\eta\big)\Big]\,+ %
}\vspace{1mm}\\ %
\displaystyle{\hspace{22mm}%
  + \sum_{n=[x_0]+2}^\infty\Phi_s\Big[\tau^{-1}\!\big(\eta-n^2\big)\Big]. %
}
\end{array}
\end{equation}

In the considered problem we need the functions at $s=1$ and $s=2$,
which for $\eta > 1$ can be represented in the form (\ref{41}), (\ref{42}). %

\end{document}